\documentclass[pre,letterpaper,twocolumn,final,superscriptaddress,floatfix]{revtex4}

\usepackage{graphicx,psfrag,amsmath,amssymb,amsfonts,bbm,latexsym,color,dcolumn,bm,mathbbol}
\usepackage{hyperref}
\usepackage{xfrac} % for sfrac{1{B}
\usepackage{ifthen}

\usepackage{amsmath,amssymb}             % AxMS Math
\usepackage[french, english]{babel}
\newcommand{\refn}[1]{Eq. (\ref{#1})}
\newcommand{\smartit}[3]{\int^{#1}_{#2} \! \mathrm{d} #3 \,}  % Raccourcis pour l'intégration
\newcommand{\pa}{\partial}

\newcommand{\GMFPT}{\left\langle \overline{t} \right\rangle }
\newcommand{\GMFPTma}{\left\langle \overline{t} \right\rangle^{\mathrm{ua}} }

\newcommand{\AMFT}{\left\langle t \right\rangle}
\newcommand{\tauopt}{\tau_{\mathrm{opt}}}
\newcommand{\tauoptma}{\tau_{\mathrm{opt}}^{\mathrm{ua}}}

\newcommand{\GMFPTopt}{\left\langle \overline{t} \right\rangle_{\mathrm{opt}}}

\newcommand{\pd}[2]{\frac{\partial #1}{\partial #2}}

\newcommand{\nm}[1]{\left|#1\right|} % Norme

\usepackage{stmaryrd}

\usepackage{enumerate}

\newcommand{\cverbose}{1}
\begin{document}

\graphicspath{{Figures/}}

\title{Optimal search strategies of run-and-tumble walks}

\author{Jean-Fran\c cois Rupprecht}
\affiliation{Sorbonne Universit\'es, UPMC Univ Paris 06, UMR 7600, Laboratoire de Physique Th\'eorique de la Mati\`ere Condens\'ee, 4 Place Jussieu, Paris (France).}
\affiliation{Mechanobiology Institute, National University of Singapore, 5A Engineering Drive 1, 117411 (Singapore).}
\email{mbijr@nus.edu.sg}

\author{Olivier B\'enichou}
\affiliation{Sorbonne Universit\'es, UPMC Univ Paris 06, UMR 7600, Laboratoire de Physique Th\'eorique de la Mati\`ere Condens\'ee, 4 Place Jussieu, Paris (France).}

\author{Raphael Voituriez}
\affiliation{Sorbonne Universit\'es, UPMC Univ Paris 06, UMR 7600, Laboratoire de Physique Th\'eorique de la Mati\`ere Condens\'ee, 4 Place Jussieu, Paris (France).}
\affiliation{Sorbonne Universit\'es, UPMC Univ Paris 06, Laboratoire Jean Perrin, UMR 8237 CNRS /UPMC, 4 Place Jussieu, Paris (France).}

%\email{voiturie@lptmc.jussieu.fr}

\begin{abstract}
The run-and-tumble walk, consisting in randomly reoriented ballistic excursions, models phenomena ranging from gas  kinetics to bacteria motility. We evaluate the mean time required for this walk to find a fixed target within a 2D or 3D spherical confinement. We find that the mean search time admits a minimum as a function of the mean run duration for various types of boundary conditions and run duration distributions (exponential, power-law, deterministic). Our result stands in sharp contrast to the pure ballistic motion, which is predicted to be the optimal search strategy in the case of Poisson distributed targets.
\end{abstract}

% Keyword: Search process, Optimization, Many-body Physics, Knudsen limit

\date{\today}

\ifthenelse{ \cverbose > 1}{}{\maketitle}

%Most organisms exhibit persistent motion, whether  as bacteria \cite{Berg1976} or larger animals.
\section{Introduction}
Run-and-tumble walks (RTW) are usually defined as  randomly reoriented ballistic excursions -- or runs -- of constant speed in ${\mathbb R}^d$. Historically, such processes where introduced to model the dynamics of particles in scattering environments, such  as the motion of atoms in a gas \cite{Einwohner1968, Knudsen1950} or electrons in a metal \cite{Martens2012,Moll2016,Zaanen2016}, as well as the scattering of light or neutrons  \cite{Zoia:2011,Zoia:2011a}. RTWs raised a renewed interest more recently as it was realized that they provide a minimal model of self-propelled particle dynamics \cite{Romanczuk:2012fk}, notoriously exemplified by swimming  bacteria  \cite{Berg2004}. Their range of applications has since then been extended from the microscopic scale, as is the case of molecular motors \cite{Julicher1997}, to the cellular scale \cite{Heuze:2013uq} or the macroscopic scale in the case of animal behavior \cite{Benichou:2011b, Shlesinger2009}. 

In the context of living agents, %self-propulsion comes at an energy  cost, and 
it is usually hypothesized  that motion is aimed at fulfilling a biological function. A prototypical example of such function is the search for targets -- as is the case for immune system cells looking for pathogens, sperm cells searching for an ovocyte, or bacteria or larger organisms foraging for food \cite{Yang2015, Meerson2015a, Benichou:2011b, Shlesinger2009}. In these examples, the optimization of the search efficiency, which appears as essential,  amounts to minimizing the search time for targets.  In this context, L\'evy walks -- RTWs with power law distributed run lengths -- have been repeatedly invoked as  optimal foraging strategies \cite{Rosa1999}. This optimality however holds only
in the particular case where the targets are distributed in space according to a Poisson law, and are in addition assumed to regenerate at the same location after a finite time. In the opposite case of destructive search, where each target can be found only once, the search is optimized by the trivial ballistic excursion. These restrictive conditions have weakened the robustness of L\'evy strategies and made their applicability to field data rather disputed over the years \cite{Edwards:2007,Benhamou:2007fk,James:2011,Palyulin2014}. In this context, assessing the search efficiency of general RTWs -- and in particular L\'evy walks -- in the basic setting of a single target in a finite domain appears as a very important issue that has remained unsettled so far: this is the goal of this article.

From a technical point of view,  a natural observable to quantify the search efficiency is  the mean first-passage time (MFPT) to a target. In the case of a single target in the interior of a bounded domain -- or equivalently  regularly spaced targets in infinite space --, the full FPT statistics, and in particular its mean, have been derived asymptotically for Markovian scale invariant random walks \cite{Condamin:2007zl,BenichouO.:2010a,Benichou2014} (see also \cite{Guerin2012} for an extension to non-Markovian processes). These results have been be extended to the narrow-escape case, in which the target is embedded in the boundary of the confining domain \cite{Singer:2006a, Singer2006,Schuss2007, Isaacson2013a, Rupprecht2014}.

While these results apply in particular to Brownian particles that are subject to thermal fluctuations, most of models of self-propelled particles, and in particular RTWs, are not scale invariant. These walks exhibit a cross-over from a short time ballistic scaling to a long time diffusive scaling. Despite their extensive analysis in infinite space \cite{Gilbert:2011}, the study of the FPT statistics of RTWs in confinement requires new approaches. First steps in this direction were made in Refs. \cite{Blanco:2003a, Benichou:2005a, Benichou2006,Benichou:2007}. More recently, the MFPT for lattice persistent random walks has been studied in Ref. \cite{Tejedor:2012ly}.  This however does not solve the case of RTWs, which cannot be defined as a  continuous limit of lattice walks -- a striking difference being that the MFPT for lattice walks can be controlled by infinitely long periodic orbits \cite{Tejedor:2012ly}, which do not exist for general domain shapes or boundary conditions in continuous space. Some previous work addressed the search optimization of RTWs in terms of their mean run time. First, Ref. \cite{Campos2015a} considers the optimal search problem by mortal RTWs, which may die at any time according to an exponential decay law. There, the optimal reorientation time maximizes the probability that the target is found before the death of the searcher. We also mention that Ref. \cite{Romanczuk2015} considers a directed RTW search where, on average, tumbles reorient the searcher in the direction of the target.  Due to this bias in the directions of reorientation,  the mean search time remains finite even in the absence of a confining domain. An optimal reorientation time appears when a waiting time is associated to each tumbling events. However, the origin of such optimum cannot apply to the problem considered here (see the concluding discussion of Sec. \ref{sec:exponential}).

In this paper, we first consider the case of a RTW with exponentially distributed run lengths that is confined within a spherical domain of radius $b$. We derive an accurate analytical approximate of the MFPT to a centered target of size $a$.  Beyond the single target problem, this setting efficiently approximates the destructive search of infinitely many regularly spaced targets with concentration $\rho = 1/(\pi b^2)$ (2D) or $\rho = 3/(4 \pi b^3)$ (3D). We show that the MFPT is minimized for a  mean run duration  $\tau =\tauopt \sim \gamma b/\log(b/a)$, where $\gamma$ is a constant that depends both on the spatial dimension and the choice of boundary conditions. However, the existence of the minimum is independent of the specific nature of the boundary condition, which pleads for its robustness. In addition, we show that similar optima exist for deterministic run durations (with respect to the run duration $\tau$) and for power law distributed run durations (with respect to the scale parameter $\tau$ of the power law), i.e. L\'evy walks. In sharp contrast with the case of Poisson distributed targets, our results highlight the crucial role of the spatial distribution of targets in search problems, and put forward the robustness  of optimal RTWs for single target search problems.

\section{Exponentially distributed runs} \label{sec:exponential}

\begin{figure}[t!]
\ifthenelse{ \cverbose > 2}{}
{
  \includegraphics[width=7.00cm]{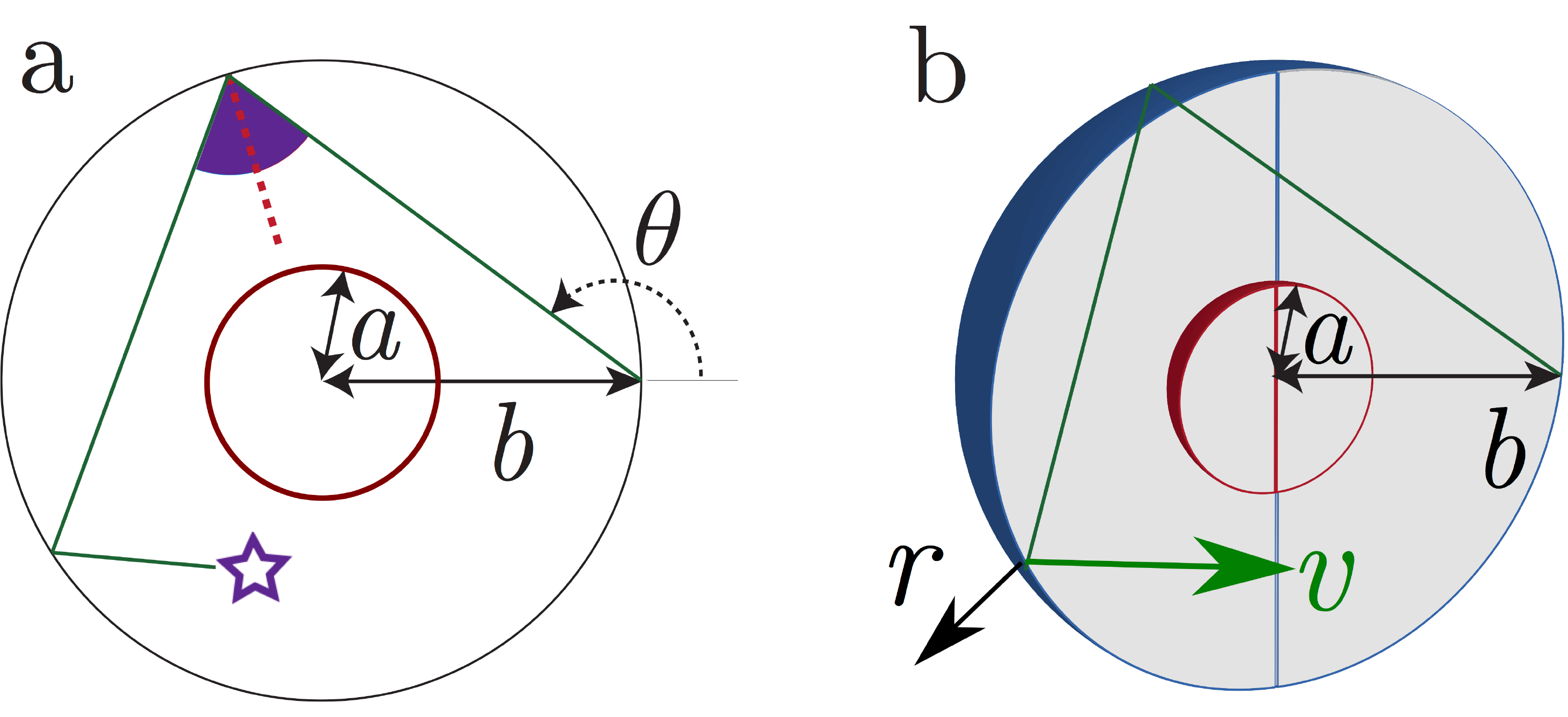}
}
  \caption{Scheme of a run-and-tumble walk. (a) In 2D, with
 specular confinement at the radius $b$, and a single tumble (star). Mind that a ballistic run which is not initially directed towards the target will not cross it at any ulterior time. (b) In 3D. Due to the spherical symmetry, the search time does not depend on the azimuthal angle $\phi$ between the velocity vector $v$ and the position vector $r$.}
\label{Fig1}
\end{figure}

We first assume that the duration of each run is exponentially distributed with mean $\tau$, which we call persistence time. We denote by $\theta$  the angle between the position vector ${\bf r}$ of the searcher and its velocity vector ${\bf v}$, the origin being the target center (see Fig. \ref{Fig1}). At each tumble, a new direction $\theta$  is set according to a uniform distribution $\mu(\theta) = 1/\pi$ in 2D and $\mu(\theta) = \sin(\theta)/2$ in 3D.
In 3D, the angle $\theta$ refers to the elevation angle of ${\bf v}$ with respect to ${\bf r}$. For symmetry reason the azimuthal angle $\phi$ can be reset to $0$ after each tumble event; hence the 3D search problem with isotropic reorientations corresponds to a 2D search problem with non-isotropic reorientations characterized by $\mu(\theta) = \sin (\theta)/2$.  

The MFPT for a searcher starting at a distance $r$ from the target center with a velocity angle $\theta$ is denoted $t(r, \theta)$. We define the averaged MFPT over all possible starting angles as $\AMFT = \int^{\pi}_{0} \! \mathrm{d}{\mu(\theta)} \, t(r, \theta)$; we will be mainly interested in the global MFPT (GMFPT), defined as the spatial average: $\GMFPT = \int^{b}_{a} \! \mathrm{d}{\nu(r)} \, \left\langle t(r) \right\rangle$, where $\nu(r)$ is the uniform spatial measure which reads $\nu(r) =d r^{d-1}/(b^d-a^d)$ in terms of the spatial dimension $d$. The MFPT can be shown to satisfy the following integro-differential backward equation \cite{Gardiner2009}:
\begin{align} \label{eq:integro}
v \cos \theta \pa_r  t(r, \theta) + \frac{1}{\tau} \int^{\pi}_{0} \! \mathrm{d}\phi \, \left\lbrace t(r, \phi)  - t(r, \theta) \right\rbrace  = -1,
\end{align}
which admits a unique solution when completed by boundary conditions. First, $t(r = a,\theta) = 0$ for $\theta \in \left[\pi/2, \pi\right]$ as the process is stopped as soon as the searcher crosses the target. Secondly, we will consider three types of confining boundary at $r = b$: (i) the specular boundary, which reflects the incoming searcher according to Descartes law, as a mirror would reflect a ray of light; (ii) the scattering boundary, which scatters the searcher in a random direction, with reinitialization of the run duration; (iii) the stubborn boundary, for which  the searcher waits for  the next favorable reorientation to leave the boundary. In the diffusive limit $\tau \ll a/v$, we expect to have effectively $\pa_r \langle t(r)\rangle = 0$ at $r=b$ for any choice of boundary condition. However, for arbitrary values of $\tau$, defining the appropriate boundary condition at $r=b$  of the MFPT is a challenging problem. 

Analytical solutions of \refn{eq:integro} are only available in 1D geometries, for  which the orientation angle is either $\theta = 0$ or $\theta = \pi$. Following Refs. \cite{Gardiner2009, Weiss1984}, we find that the mean search time in the presence of a specular boundary condition reads $\GMFPT =  ((a-b)^2(3 D-(a-b)v))/(3 b D v)$, where $D = v^2\tau$. Mind that the latter expression predicts that the GMFPT is a monotonically decreasing function of $\tau$, in sharp contrast with the main results of this paper concerning the 2D and 3D geometries. As no such exact solutions are available in these higher dimensions, we resort below to an approximate scheme that interpolates the exact asymptotics  obtained in both limits $\tau \ll a/v$ and $b/v \ll \tau$. 

\subsection{Diffusive limit} \label{sec:diffusivelimit}
In the diffusive limit $\tau \ll a/v$, the MFPT is independent both of the direction of the initial velocity $\theta$ and of the nature of the confining boundary (e.g. specular, scattering or stubborn). Following  \cite{Benichou:2011b}, it can be shown that in this limit the MFPT  satisfies the equation: $D \Delta \left<  t(r) \right>=-1$ where $D = v^2\tau/d$. In  the large volume  limit $a\ll b$, it is straightforward to obtain that \cite{Redner:2001a}
\ifthenelse{ \cverbose > 2}{}
{
\begin{align} \label{eq:taularge_diffusive}
\GMFPT \underset{\tau  \ll a/v}{\sim}  \overline{t}^{(1)} = \frac{b^2}{v^2 \tau} \ln\left(\frac{b}{a}\right),
\end{align}
}
in 2D, and  $\GMFPT \sim \overline{t}^{(1)} = \pi b^3/(3 a v^2 \tau)$ in 3D. We emphasize that in the diffusive limit, the search time is a decreasing function of the persistence time, for any type of confining boundary (specular, scattering or stubborn).

\subsection{Ballistic limit and uniform approximation} \label{sec:specular_exponential}
In the limit $\tau \gg b/v$, the behavior of the search time crucially depends on the  boundary condition at $r=b$. We derive the expression of the optimal reorientation time for three particular types of boundary condition.

\subsubsection{Specular boundary condition} We first consider the case of a specular boundary that reflects incoming searchers according to Descartes law. To evaluate the search time in this regime, we follow an approach introduced in Ref. \cite{Romanczuk2015}, which we call uniform approximation. This approach also shares similarity with the method presented in Ref. \cite{Isaacson2013a}. We first write the probability of hitting the target within a single run after a tumble event as
\ifthenelse{ \cverbose > 2}{}
{
\begin{align} \label{eq:ph_refl}
p_h = \int^{b}_{a} \mathrm{d}r \, p_r(r)  p_{\mathrm{dir}}(r)  p_{\mathrm{reach}}(r),
\end{align}
}
where:
(i)  $ p_r(r)$ is the probability distribution of the position at the previous reorientation event. For $r\gg a$, it can be assumed that $p_r$ is uniform over the domain. 
%However, this approximation should fail in the diffusion limit, due to the depletion caused by the target.
%We assume that $p_r$ has relaxed to a steady state, which, in the limit $a/b \ll 1$, is the uniform distribution $\mathrm{d} p_r(r) = \mathrm{d}\mu(r)$,  
(ii)  $p_{\mathrm{dir}}(r)$ is the probability of choosing a direction that crosses the target (either directly or after a single reflection at $r=b$). In the case of a specular confining boundary, $p_{\mathrm{dir}}(r) = 2 \cdot a/(\pi  r)$ (2D) and $p_{\mathrm{dir}}(r) = 2 \cdot a^2/(4 r^2)$ (3D) at first order in $a/r \ll 1$. The factor $2$ in the latter two expressions is due to trajectories that cross the target after a single reflection on the boundary $r=b$.
(iii)  $p_{\mathrm{reach}}(r)$ is the probability of performing a sufficiently long run to reach the target, which for exponentially distributed runs reads:
\begin{align} \label{eq:preach}
p_{\mathrm{reach}}(r) = \left( e^{-r/(v\tau)}+ e^{-(2b-r)/(v\tau)} \right)/2.
\end{align}
Partitioning over the first run, one writes $\GMFPT = p_h \tau_f + (1-p_h) (\GMFPT+\tau)$, where $\tau_f$ is the mean duration of a successful run.   In the joint ballistic and small target limit ($a/b \ll 1$), we find that $p_h \ll 1$, hence that $\GMFPT \sim \tau/p_h$ and
\begin{align} \label{eq:mfpt_tau_large_specular}
\GMFPT \underset{b/v \ll \tau}{\sim} \overline{t}^{(2)} \equiv \frac{\pi  b^2}{2 v a} \, \frac{1}{\left(1-e^{-\frac{2 b}{\tau v}}\right)}  \underset{b/v \ll \tau}{\sim} \tau \, \frac{\pi}{4} \frac{b}{a},
\end{align}
in 2D, while in 3D,
\begin{align} 
\GMFPT \underset{b/v \ll \tau}{\sim} \overline{t}^{(2)} \equiv  \frac{4 b^3}{3 a^2 v} \frac{1}{\left(1-e^{-\frac{2 b}{\tau v}}\right)} \underset{b/v \ll \tau}{\sim} \tau \frac{2}{3} \frac{b^2}{a^2}.
\end{align}
The GMFPT is therefore an increasing function of $\tau$ in the limit $\tau \gg b/v$ that diverges in the limit $\tau = \infty$. Indeed, a ballistic particle that is not initiated in the direction of the target will never cross the target. 
Together with the above analysis of the diffusive regime, this shows that the GMFPT can be minimized as a function of $\tau$.

\begin{figure}[t!]
  \includegraphics[width=8.25cm]{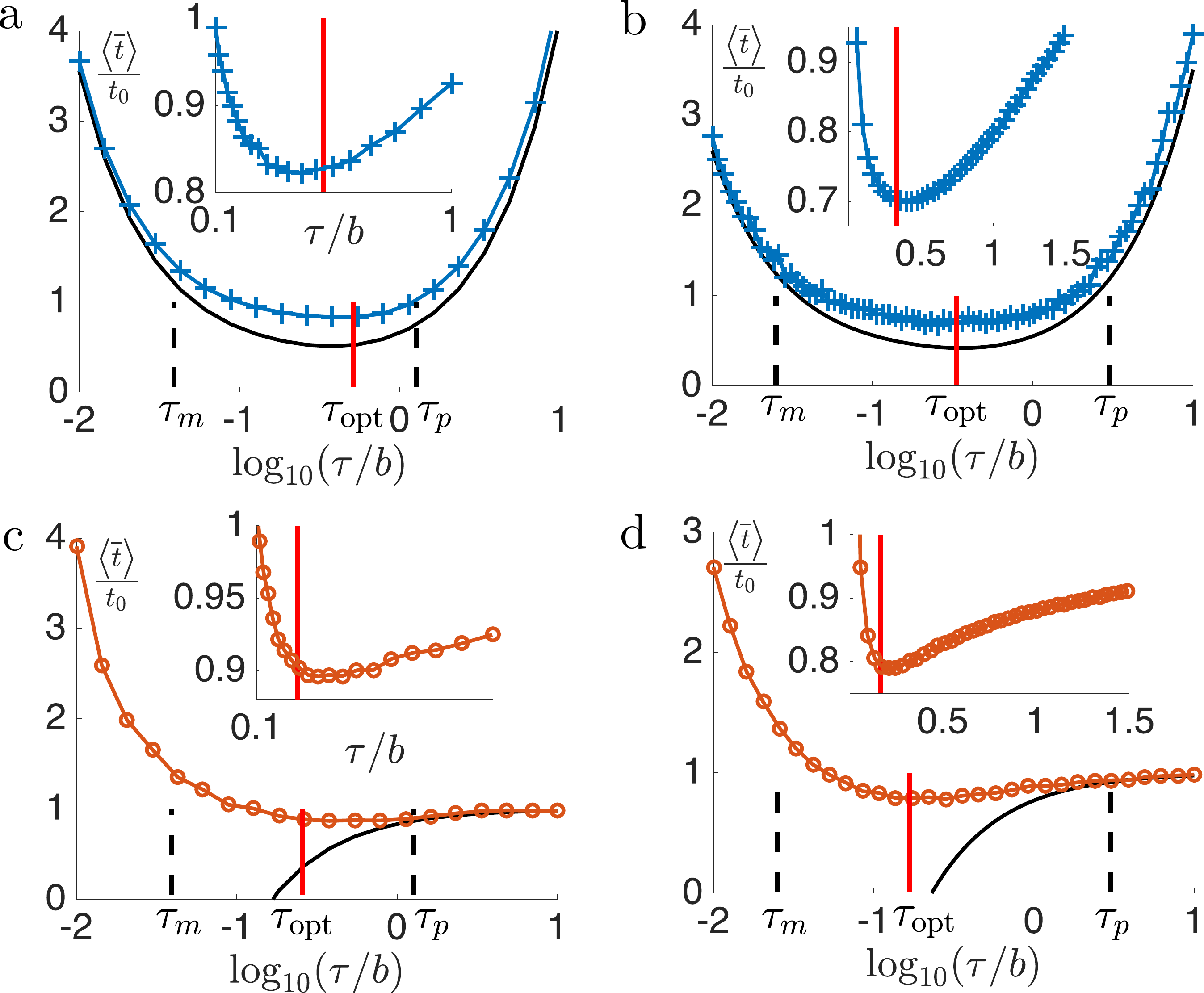}
  \caption{(Color online) Search time (GMFPT) to a disk (2D) or a sphere (3D) of radius $a$ as a function of the persistence time $\tau$: (a, c) in 2D, with a confinement radius $b = 50 \, a$, (b, d) in 3D, with $b/a= 20$, obtained by simulations with a confining boundary that is (a, b, blue crosses) specular or (c, d, orange circles) scattering ; (a, b, solid black line) analytical prediction expression from \refn{eq:matched_asymptotic}; (c, d, black dashed circles) asymptotic relations from Eqs. (\ref{eq:tauf_diffusive2D}) and (\ref{eq:tauf_diffusive3D}). The search time is close to $1$ over the predicted plateau range $\left[\tau_m/b, \tau_p/b\right]$ ($\tau_m$ and $\tau_p$ (black vertical dashed lines). (Insets) Search time at optimality:  (red solid vertical lines) optimal persistence time from (a, b) \refn{eq:tauopt} and (c, d) \refn{eq:tauoptma_scattering}. The search time is scaled by $t_0 = 2 (b/v) (b/a)^{d-1}$.}
\label{Fig2}
\end{figure}

\begin{figure}[t!]
  \includegraphics[width=8.50cm]{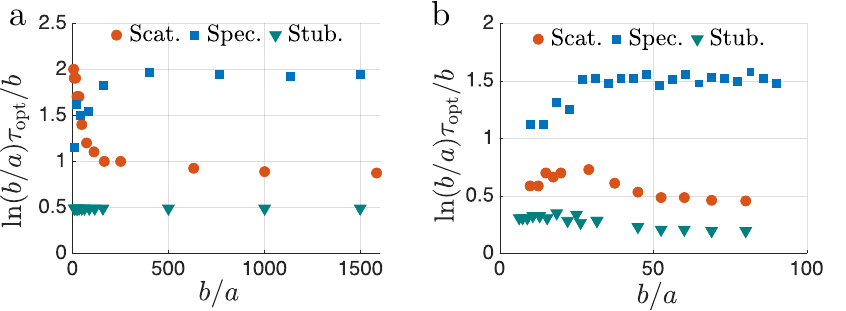} 
  \caption{(Color online) Scaling of the optimal reorientation time: convergence of the ratio $\gamma = \log(b/a) v \tau_{\mathrm{opt}}/b$ as a function of the domain size $b$ (a) in 2D and (b) in 3D. The corresponding boundary condition is (orange circle) scattering; (blue square) specular; (green triangle) stubborn. }
\label{Fig3}
\end{figure}

We now verify that an accurate uniform approximate of the GMFPT is given by 
\begin{align} \label{eq:matched_asymptotic}
\GMFPTma = \overline{t}^{(1)} + \overline{t}^{(2)}.
\end{align}
The fact that $\overline{t}^{(2)} \to 0$ in the diffusive limit and $\overline{t}^{(1)} \to 0$ in the ballistic limit shows that this approximation is accurate in both limit regimes; it was further checked numerically that the agreement is good for all values of $\tau$ (see Fig. \ref{Fig2}a,b). Importantly, this approximate solution makes it possible to analyze analytically  the minimization of the GMFPT. The condition of minimization of expression (\ref{eq:matched_asymptotic}) leads to:
\begin{align}\label{eq:uniformapproximate_specular}
\frac{\pi  b^3 X}{a (1-X)^2}-b^2 \log \left(\frac{b}{a}\right)=0,
\end{align}
in terms of $X = \exp(-2 b/(v\tau))$. In the large volume limit $a \ll b$, the condition defined in \refn{eq:uniformapproximate_specular} leads to the following expression for the optimal reorientation time:
\begin{align}\label{eq:tauoptma}
\tauoptma \underset{a \ll b}{\sim} 2 \frac{b}{v \ln(b/a)},
\end{align}
in both 2D and 3D. Analysis of the simulations confirm the scaling of  \refn{eq:tauoptma}, as we find that:
\begin{align}\label{eq:tauopt}
\tauopt \underset{a \ll b}{\sim} \gamma^{\mathrm{spec}}_d \frac{b}{v \ln(b/a)},
\end{align}
where $\gamma^{\mathrm{spec}}_d$ is a constant that only depend on the space dimension $d$: $\gamma^{\mathrm{spec}}_2 = 1.95 \pm 0.02$ in 2D and $\gamma^{\mathrm{spec}}_3 = 1.5 \pm 0.02$ in 3D (see Fig. \ref{Fig3}). Hence, the uniform approximate provides the $b/\log(b/a)$-scaling of the optimal reorientation time, but it estimates only approximately the prefactor $\gamma^{\mathrm{spec}}_d$. 

\begin{figure}[t!]
  \includegraphics[width=8.50cm]{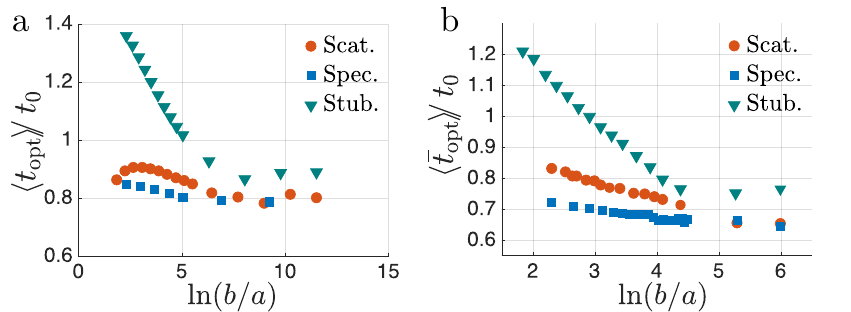} 
  \caption{(Color online) Scaling of the optimal search time: convergence of the ratio $\GMFPTopt/t_0$, with $t_0 = 2 (b/v) (b/a)^{d-1}$, as a function of the domain size $b$ (a) in 2D and (b) in 3D. The corresponding boundary condition is (orange circle) scattering; (blue square) specular; (green triangle) stubborn. For large values of the confining domain ($\ln(b/a)> 6$ in a; $\ln(b/a)> 4$ in b) we estimate the search time at the value of the reorientation time indicated by Eqs. (\ref{eq:tauopt}--\ref{eq:tauoptma_scattering}--\ref{eq:tauoptma_stubborn_simulation}). }
\label{Fig4}
\end{figure}

At the optimal reorientation time, the minimal mean search time scales as
\begin{align}\label{eq:searchopt}
\GMFPTopt \underset{a \ll b}{\sim} 2 \xi^{\mathrm{spec}}_d (b/v) (b/a)^{d-1},
\end{align}
where $\xi^{\mathrm{spec}}_d$ are constants that read $\xi^{\mathrm{spec}}_2 = 0.825 \pm 0.01$ in 2D and $\xi^{\mathrm{spec}}_3 =  0.64 \pm 0.02$ in 3D. The convergence of the ratio $\GMFPTopt/(b/a)^{d-1}$ is represented in Fig. \ref{Fig4}. 

We find that the mean search time exhibits a large plateau in $\tau$ around $\tau_{\mathrm{opt}}$ in which it does not significantly differ from $\GMFPTopt$ (see Fig. \ref{Fig2}). We estimate the lower bound of the plateau $\tau_{m}$ as the cross-over reorientation time from the diffusive limit to the optimal regime, i.e. as the solution of the equation: $ \overline{t}^{(1)}(\tau_{m}) =  \overline{t}_{\mathrm{opt}}$. We obtain $\tau_{m} = a/(2 v) \ln\left(b/a\right)$ in 2D, and $\tau_{m} = \pi a/(6 v)$ in 3D.  Similarly, for the upper bound of the plateau $\tau_{p}$, the condition $ \overline{t}^{(2)}(\tau_{m}) =  \overline{t}_{\mathrm{opt}}$ leads to $\tau_{p} = (8 b)/(\pi v)$ (2D) and $\tau_{p} = (3 b)/(v)$ (3D). Mind that the length of the interval $\tau_{p}-\tau_{m}$ is an increasing function of $b$: the search time is all the more flat around its optimum as the size of the confinement $b$ is increased. Therefore, the statistical uncertainty on the value of the optimal reorientation time increases significantly with the size of the confinement radius. In Fig. \ref{Fig3}, the search time -- which is computed for each value of $b$ and $\tau$ -- is averaged over a larger number of trajectories: $10^{7}$ in 2D and $10^{5}$ in 3D.

%Only the specular boundary condition was considered in \cite{Tejedor:2012ly}, in which the optimum is interpreted as a compromise between limit behavior with diverging search time. Here, we show that optimum is not specific to the specular nature of boundary condition that makes the ballistic search time diverge, but rather is an intrinsic property of the run-and-tumble process. 
%Here, we show that the optimum is robust to a change of boundary condition.

\subsubsection{Scattering boundary} \label{sec:existence}  We now consider the case of a scattering boundary -- at which the searcher is scattered in a random direction. This condition corresponds to a widely used assumption about gas scattering on weakly adsorbing surfaces, known as Knudsen cosine law \cite{Feres}. It can also be used to model the behavior of bacteria at the encounter of a boundary \cite{Tailleur2008}. We mention that Ref. \cite{Yang2015} considered a similar problem: the narrow escape of a ballistic particle confined by an otherwise scattering boundary.

We introduce the quantity $p_s$ as the probability that the target is found between two successive scattering events at the boundary. We introduce its large $\tau$ expansion  by writing $ p_s = p^{(0)}_s  + p^{(1)}_s/\tau+{\cal O}(1/\tau^2)$ at second order. The quantities $p^{(n)}_s/\tau^n$ correspond to the probability for the searcher to perform $n$ tumble events before hitting the target and before visiting the boundary. At first order in $a \gg b$, the probability $p^{(0)}_s$ to reach the target within a single run from the boundary reads in 2D:
\begin{align} \label{eq:ph0}
p^{(0)}_s = \frac{2}{\pi} \arcsin\left(\frac{a}{b} \right) \underset{a \ll b}{\sim} \frac{2a}{b\pi},
\end{align}
while $p^{(0)}_s = a^2/(2 b^2)$ in 3D.
% -- indeed the average cord length, denoted $v \tau _f$, is:  $v \tau _f = 4 b/(v\pi)$ (2D) and $\tau _f =  b/v$ (3D). 

The next order probability $p^{(1)}_s/\tau$ can be calculated as a sum of two terms: $p^{(1)}_s/\tau = p^{(1)}_{+}/\tau - p^{(0)}_s/\tau$. The negative contribution ($- p^{(0)}_s/\tau$) corresponds to the probability that a tumble occurs on a run that is initiated towards the target at the boundary. The positive contribution ($p^{(1)}_{+}$) corresponds to the probability that a tumble provides the direction to the target. The latter quantity can be expressed by the following integral:
\begin{align} \label{eq:ph1}
p^{(1)}_{+}  = 2 \smartit{\pi-\theta_c}{\pi/2}{\mu(\theta)} \smartit{-2b\cos(\theta)}{0}{t} p_{\mathrm{dir}}(r(t)) p_{\mathrm{reach}},
\end{align}
where (i) $p_{\mathrm{reach}} = 1$ at first order in $\tau$; (ii) $\theta_c = \arcsin(a/b)$ is the angular aperture of the target as seen from the boundary; (iii) $r(t) = \sqrt{v^2 t^2 + 2 v b t \cos \theta + b^2}$ is the position along the cord at the time $t$ after departure from the boundary; (iv) $p_{\mathrm{dir}}(r)$ is the probability of choosing a direction that crosses the target after a reorientation event at the radius $r$. At first order in $a/r \ll 1$, one finds $p_{\mathrm{dir}}(r) = a/(\pi  r)$ (2D) or $p_{\mathrm{dir}}(r) = a^2/(4 r^2)$ (3D).  We perform the change of variable $u = b t/v$ in \refn{eq:ph1} to obtain the scaling relation $p^{(1)}_s = (a/v) (a/b)^{d-2} \chi_d$ where $\chi_d$ equals:
\begin{align} \label{eq:chid}
2 \smartit{\pi}{\pi/2}{\mu(\theta)} \smartit{-2 \cos(\theta)}{0}{u} p_{\mathrm{dir}}(\sqrt{u^2 + 2 u \cos \theta + 1}). 
\end{align}
Numerical integration leads to the values $\chi_2 \approx 0.74$ in 2D and $\chi_3 \approx 0.61$ in 3D.  
%We define the mean cord length between two point on the boundary $r = b$, which reads $\tau_c = (4/\pi) b$ in 2D and $\tau_c = b$ in 3D. 
%From this expression, we justify the scaling of $p^{(1)}_{+}$ as we expect a relation of the type $p^{(1)}_{+} \propto \tau_c \, p^{(0)}_s$.  

We next introduce the time $\tau_c$ corresponding to the mean length of cords between two points on the boundary $r = b$, which reads $\tau_c = (4/\pi) b$ in 2D and $\tau_c = b$ in 3D. In the limit $b\gg a$, the GMFPT is then given by   $\GMFPT\sim \tau_c/p_s$. Using the above expression for $p_s$, we obtain:
\begin{align} \label{eq:tauf_diffusive2D}
\GMFPT = \frac{2 b^2}{v a} - (\pi \chi_2 - 2) \frac{b^3}{v^2 a}  \cdot \tau^{-1}+ o(\tau^{-1}),
\end{align}
in 2D, while  in 3D 
\begin{align} \label{eq:tauf_diffusive3D}
\GMFPT = \frac{2 b^3}{v a^2} - (2 \chi_3 - 1) \frac{2 b^4}{v^3 a^2}  \cdot \tau^{-1}  + o(\tau^{-1}).
\end{align}
Mind that $\pi \chi_2 - 2 \approx 0.17 >0$ and similarly $2 \chi_3 - 1 \approx 0.23  >0$. We verify the predictions of Eqs. (\ref{eq:tauf_diffusive2D}--\ref{eq:tauf_diffusive3D}) in Figs \ref{Fig2}c,d. 

We conclude that, in the ballistic limit, increasing the rate of tumble $1/\tau$ necessarily decreases the search time as compared to the pure ballistic motion. Our intuitive interpretation for this behavior is that tumbles may occur close to the target, where the probability for a reorientation in the direction of the target is high. The asymptotic relations Eqs. (\ref{eq:tauf_diffusive2D}) and (\ref{eq:tauf_diffusive3D}) thus prove the existence of an optimum when the confining boundary is scattering in 2D and 3D, respectively.

For this case of scattering boundary conditions, the uniform approximation cannot be used to determine the minimum of the MFPT, as the MFPT converges to a finite value in the limit $\tau\to \infty$. However, analysis of numerical simulations confirm a similar scaling to \refn{eq:matched_asymptotic} for the optimal reorientation time:
\begin{align}\label{eq:tauoptma_scattering}
\tauopt \underset{a \ll b}{\sim} \gamma^{\mathrm{scat}}_{d} \frac{b}{v \ln(b/a)},
\end{align}
where $\gamma^{\mathrm{scat}}_{2}  = 0.87 \pm 0.2$ and  $\gamma^{\mathrm{scat}}_{3}  = 0.45 \pm 0.02$ (see Fig. \ref{Fig3}). At the optimal reorientation time, the minimal mean search time scales as
\begin{align}\label{eq:searchopt_scat}
\GMFPTopt \underset{a \ll b}{\sim} 2 \xi^{\mathrm{scat}}_d (b/v) (b/a)^{d-1},
\end{align}
where $\xi^{\mathrm{scat}}_2 = 0.88 \pm 0.01$ in 2D and $\xi^{\mathrm{scat}}_3 =  0.65 \pm 0.02$ in 3D.  The convergence of the ratio $\GMFPTopt/(b/a)^{d-1}$ is represented in Fig. \ref{Fig4}. 

We point out that $\xi^{\mathrm{scat}}_d = \GMFPTopt/t_0$ corresponds to a measure of the search efficiency at optimality, where we recall that
 $t_0 = 2 (b/v) (b/a)^{d-1}$ corresponds to the GMFPT in the ballistic limit. As $\xi^{\mathrm{spec}}_d < \xi^{\mathrm{scat}}_d$, we conclude that the optimal search process is faster for a specular boundary than for a scattering one. This can be interpreted by a geometrical factor: in the ballistic limit, the probability to select a direction that leads to the target $p_{\mathrm{dir}}(r)$ is two-times higher for a specular boundary than for a scattering boundary.

\begin{figure}[t!]
  \includegraphics[width=8.50cm]{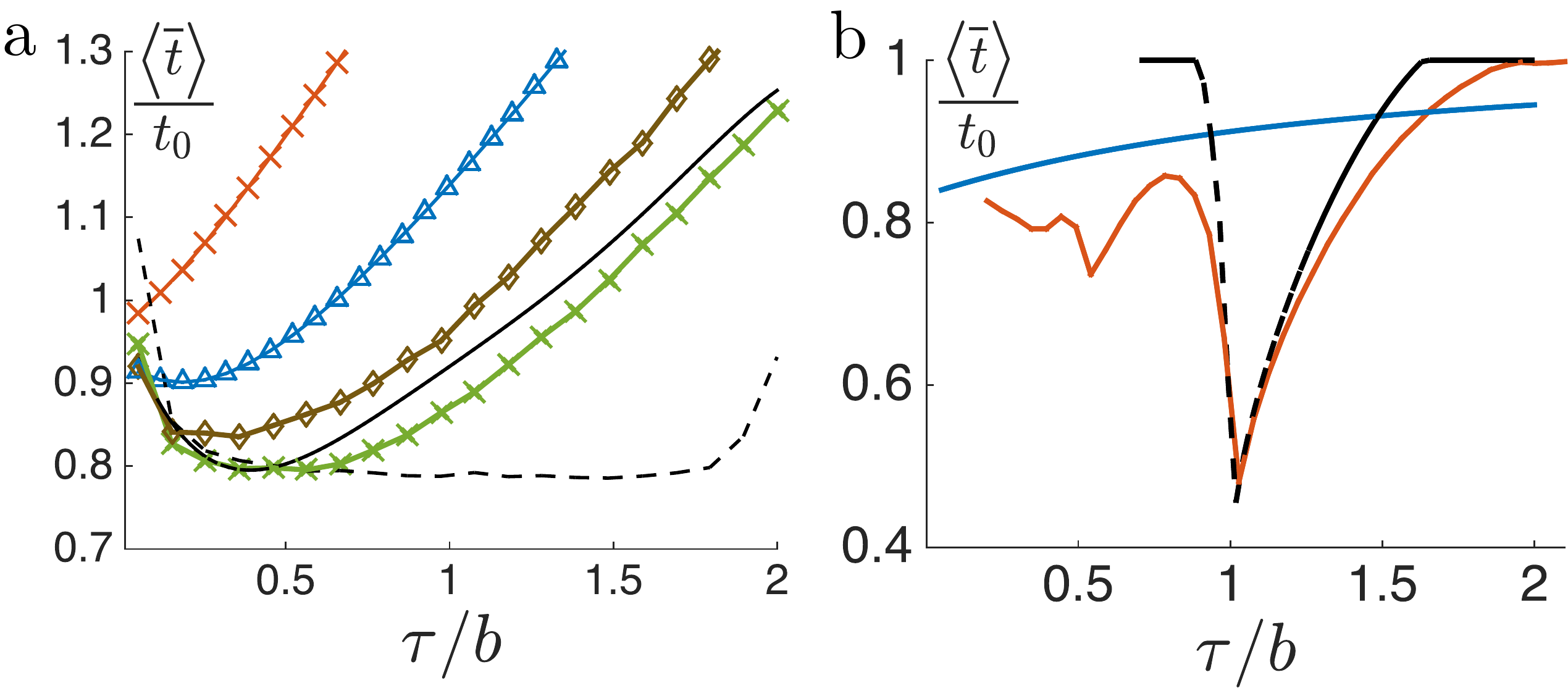}
  \caption{(Color online) (a) Search time of a 2D disk with confinement at $b/a =200$ with specular confinement, for (solid colored lines with symbols) L\'{e}vy walks with the following values of $\alpha$: $1.4$, $1.6$, $1.8$, and $2$ (from top to bottom) ; (solid black line) exponentially-distributed step walk ; (dashed black line)
regular step walk. (b) Search time of a 2D disk with confinement at $b/a =200$ for a regular-step walk with scattering confinement: simulations  (solid orange line) indicate a sharp minimum at $\tau = b$, as predicted by our approximate expression (dashed black lines), which outperforms the exponential walk (solid upper blue line). }
  % Data generated with: phase_generation_circle_2
  % Plots obtained with: compare_distribution
\label{Fig5}
\end{figure}

\subsubsection{Stubborn boundary}

We call stubborn condition the case in which the searcher sticks upon arrival at the external boundary, and remains at the boundary for the remaining time before the next reorientation towards the interior of the domain. This situation corresponds to the case of  fully inelastic collision, which is often used to describe overdamped self-propelled particles in confinement \cite{Ghosh2013}. Indeed, a fully inelastic collision between the searcher and the boundary results in the  sliding motion of the searcher along the boundary; due to the complete spherical symmetry of the problem, this sliding motion condition is equivalent to the particle being stuck in place.

Under this choice of boundary condition, we expect the search time to diverge in the ballistic limit $b/v \ll \tau$. More precisely, we expect that  $\GMFPT\sim \tau/p^{(0)}_s$ where $p^{(0)}_s$ is the probability $p^{(0)}_s$ to reach the target in a single run from the boundary, following the definition of \refn{eq:ph0}. In the large volume limit $a/b \ll 1$, the latter expression for the GMFPT reads 
\begin{align} \label{eq:mfpt_tau_large_specular}
\GMFPT \underset{b/v \ll \tau}{\sim} \overline{t}^{(2)} \equiv 2 \tau b^{2}/a^{2},
\end{align}
while $\GMFPT \sim \overline{t}^{(2)} \equiv 4 b^2 \tau/(3 a^2)$ in 3D. 

Therefore, the GMFPT is an increasing function of $\tau$ in the ballistic limit. On the other hand, we justified in Sec. \ref{sec:diffusivelimit} that the search time is a decreasing function of $\tau$ in the diffusive limit. Therefore we expect the GMFPT to be an optimizable function of the reorientation time $\tau$. 

We now seek an analytical expression for the optimal reorientation time $\tauopt$; similarly to the specular case,  we define an uniform approximate of the GMFPT $\GMFPTma$. The minimization of the uniform approximate of the GMFPT leads to the condition
\begin{align} \label{eq:uniformapproximate_stubborn}
\frac{\pi  b X}{a \left(1-X \right)^2}-2 \log \left(\frac{b}{a}\right)=0,
\end{align}
in terms of $X = \exp(-b/(v\tau))$, in both the 2D and 3D cases. We then expand the solution of \refn{eq:uniformapproximate_stubborn} in the large volume limit $a \ll b$ and we find the following expression for the optimal mean run duration (in both 2D and 3D)
\begin{align}\label{eq:tauoptma_stubborn}
\tauoptma \underset{a \ll b}{\sim} \frac{b}{v \ln(b/a)}.
\end{align}
Analysis of simulations confirm the scaling of  \refn{eq:tauoptma_stubborn} and we find that:
\begin{align}\label{eq:tauoptma_stubborn_simulation}
\tauopt \underset{a \ll b}{\sim} \gamma^{\mathrm{stub}}_d \frac{b}{v \ln(b/a)},
\end{align}
where $\gamma^{\mathrm{stub}}_2 = 0.49 \pm 0.01$ in 2D and $\gamma^{\mathrm{stub}}_3 = 0.19 \pm 0.02$ in 3D. We conclude that the uniform approximation provides the appropriate scaling of the optimal reorientation time, but not the exact value of the proportionality factor $\gamma_{\mathrm{stub}}$. At the optimal reorientation time, the minimal mean search time scales as
\begin{align}\label{eq:searchopt_stub}
\GMFPTopt \underset{a \ll b}{\sim} 2 \xi^{\mathrm{stub}}_d (b/v) (b/a)^{d-1},
\end{align}
where $\xi^{\mathrm{stub}}_2 = 0.88 \pm 0.05$ in 2D and $\xi^{\mathrm{stub}}_3 =  0.76 \pm 0.05$ in 3D. Equation \refn{eq:searchopt_stub} corresponds to the same scaling as in \refn{eq:searchopt_scat} and \refn{eq:searchopt}. The convergence of the ratio $\GMFPTopt/(b/a)^{d-1}$ appears to be relatively slow compared to the specular and scattering cases, hence the larger confidence intervals on the value of $\xi^{\mathrm{stub}}_d$ (see Fig. \ref{Fig4}).

\subsubsection{Discussion} \label{sec:one_discussion}

For the three  choices of boundary conditions discussed above, we found similar scaling expressions for the optimal reorientation time $\tauopt$ and for the optimal search time $\GMFPTopt$.  The existence of an optimal reorientation time appears to be an intrinsic property of RTWs in confinement, independently of the nature of the boundary condition.  We emphasize that, compared to pure ballistic motion, tumbles allow for reorientations close to the target. However, frequent tumble results in a redundant exploration of regions distant to the target. Therefore, our intuitive interpretation is that the existence of an optimal reorientation time results from a trade-off between an enhanced ability to explore the region near the target and an increased risk to return to regions that have already been explored.  We emphasize that the predicted scaling for the optimal search time (i.e. $\GMFPTopt \propto (b/v) (b/a)^{d-1}$) is characteristic of a non-compact process \cite{Condamin:2007zl, Benichou2014}; hence the optimal search is significantly more efficient than a diffusive search, which scales according to \refn{eq:taularge_diffusive}. 

We now compare our results to previous studies on RTWs in confinement. 

First,  Ref. \cite{Tejedor:2012ly} analyzed a discrete version of the problem considered in the present work, where the target was a single site  of a regular  lattice. Ref. \cite{Tejedor:2012ly} suggested a linear scaling  of the optimal reorientation time in terms of the confinement radius $b$: $\tauopt = \xi b/v$, with a relatively small proportionality coefficient $\xi \approx 0.10$. However, we suggest that the observed low value for $\xi$ could result from a contribution of a $1/\log(b/a)$ correction.

Secondly, we point out that Ref. \cite{Benichou2006} considers a  geometry which is identical to the one considered in the present paper. However, the choice for the detection criteria was very different in this earlier reference, since the target could be detected only during tumbling events, and not during runs. This lead to very different criteria for minimizing the search time. 

Thirdly, we emphasize that in Ref. \cite{Romanczuk2015}, the RTW is biased towards the target. The optimal reorientation time is found when a finite waiting time is associated to each tumble event.  The existence of an optimal reorientation time requires that a finite waiting time is associated to each tumble events. Therefore, we expect that the prediction for the optimal reorientation time in \cite{Romanczuk2015} (ie. $\tauopt \propto a^{1/3}$) to be different from the one presented in our paper. Nevertheless, it is worth noticing that the existence of an optimal reorientation time in \cite{Romanczuk2015} results from a cost-risk trade-off between short runs, which reduces the risk of picking a wrong direction but at the cost of frequent stopping, and long runs, which are efficient to explore space but at the risk of missing the target.

Finally, the very recent numerical study \cite{Wang2016} considers the first-passage properties of self-propelled particles which change their orientation through rotational diffusion (instead of tumbles here). In contrast to our result, Ref. \cite{Wang2016} concludes on an expression for the optimal reorientation time that is proportional to the size of the confining domain, yet with a relatively low value for the proportionality coefficient. However, the existence of an optimal search strategy appears
to be a generic property of persistent random walks

\section{L\'evy walks} We next study numerically the case of L\'evy walks, whose run durations are distributed according to a symmetric L\'evy law that is restricted to the positive axis, with an index $\alpha > 1$ and a scale parameter $\tau$. In particular, the distribution of run duration is heavy-tailed $P(t) \sim t^{-\alpha -1}$ for $t \rightarrow \infty$. Here, we focus on a 2D geometry with specular boundary conditions. For $0< \alpha < 1$, the persistence length is infinite, hence, with specular boundary, the search time is infinite. For all $\alpha  > 1$, the search time can be minimized as a function of $\tau$. The optimal value $\tauopt$ decreases when $\alpha$ is increased (see Fig. \ref{Fig5}a). In particular, the search time for the L\'evy strategy is minimized when $\alpha = 2$, i.e. when the length of the ballistic excursions has a finite second moment and the central limit applies. We point out that the choice of a L\'evy walk, even with exponents close to $\alpha = 2$, does not provide any quantitative advantage to find the target over exponential-RTW (see Fig. \ref{Fig5}a). %Therefore, our result question the optimality of L\'evy walks as foraging strategies in confined geometries.

\section{Regular walks} 

In this section we focus on the 2D geometry and we find that the regular-step walk, i.e. with reorientations at deterministic time intervals $\tau$, can outperform both L\'evy and exponential RTWs, both in the presence of a specular or a scattering confinement.

With a specular confining boundary, we find numerically that the mean search time is minimized for $\tauopt = \gamma^{\mathrm{r\lvert spec}}_2 b/v$, where $\gamma^{\mathrm{r\lvert spec}}_2 = 1.4 \pm 0.1$. The optimal search time also follows a scaling of  $\GMFPTopt \underset{a \ll b}{\sim} 2 \xi^{\mathrm{r\lvert spec}}_d (b/v) (b/a)^{d-1}$, where $\xi^{\mathrm{r\lvert spec}}_2 = 0.78 \pm 0.01$. 
% and $\gamma_3 = 0.9 \pm 0.3$ in 3D

With a scattering confining boundary, simulations indicate that the GMFPT displays a sharp minimum at $\tauopt = \gamma^{\mathrm{r\lvert scat}}_2 b/v$ with $\gamma^{\mathrm{r\lvert scat}}_2 = 1.01 \pm 0.01$ (see Fig. \ref{Fig5}b). The GMFPT also displays local minima at $\tau = b/(2v)$ and $\tau = b/(3v)$. Indeed, the values  $\tau = b/(nv)$ maximize the probability to reach the target after $n$–th reorientation events before a new scattering event at the boundary. We present an analytical description of the minimum at $\tau = b/v$ in the App. \ref{app:regular}. At optimality, the search time scales as $2 (b/v) (b/a)$, with a relatively small proportionality constant $\xi^{\mathrm{r\lvert scat}}_2 = 0.28 \pm 0.02$. 

As $\xi^{\mathrm{r\lvert spec}}_2 < \xi^{\mathrm{spec}}_2$ and $\xi^{\mathrm{r\lvert scat}}_2 < \xi^{\mathrm{scat}}_2$, we conclude that the regular RTW with constant step durations can outperform any exponential RTWs.

Similar scalings hold in the stubborn case, both for the optimal reorientation time and for the optimal search time, and we find that the proportionality constants read $\gamma^{\mathrm{r\lvert scat}}_2 = 1.01 \pm 0.01$ (similarly to the scattering case) and $\xi^{\mathrm{r\lvert stub}}_2 = 0.78 \pm 0.01$ (similarly to the specular case).

\section{Curved geometry} 
\begin{figure}[t!]
  \includegraphics[width=8.50cm]{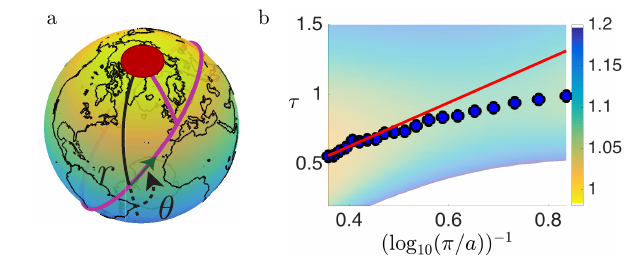}
  \caption{(Color online) Search of a cap on a sphere (a) The coordinate $r \in \left[0, \pi\right]$ represents the latitude on the sphere. The target is a spherical cap of latitude $a$ (red circle). (b) Search time normalized by $t_0 = v/(\pi a)$, in terms of the persistence time $\tau$ and of the size of the target $a$ (measured as $\log_{10}(\pi/a)^{-1}$), and optimal reorientation time $\tauopt$ obtained by (blue circles) simulations and (red line) curve $\tauopt$ from \refn{eq:tauopt_sphere}.
}
  %($10^7$ walks)
\label{Fig6}
\end{figure}

We now show that these results can be extended to other geometries. We consider a RTW with exponentially distributed runs on a sphere, which is represented in Fig. \ref{Fig6}a. The coordinate $r$ here refers to the latitude on the sphere, with  $r \in \left[0, \pi \right]$; the target consists in the North pole region with maximal latitude $a$. The implementation of simulation is detailed in the App. \ref{sec:sphere_geometry}. We now proceed to the analysis of the optimal reorientation time using the uniform approximate scheme developed above. 

In the diffusive limit, the search time reads $\GMFPT \sim (8/(v \tau)) \ln(\pi/a)$ when $a \ll 1$ \cite{Singer:2006a}. To study the ballistic limit, we define the probability $p_h$ according to \refn{eq:ph_refl}. In the limit $a \ll 1$, we expect the probability distribution of the position at reorientation events to follow the uniform distribution on the sphere, hence that $p_r(r) \sim \sin(r)/2$. The probability of choosing a direction that crosses the target, either directly or after a passing over the South pole, reads $p_{\mathrm{dir}}(r) = (2/\pi) \theta_a $, where $\theta_a $ is defined in \refn{eq:si_thetaa}. Last, the probability of performing a sufficiently long run to reach the target reads: $p_{\mathrm{reach}}(r) = \left( e^{-r/\tau}+ e^{-(2 \pi-r)/\tau} \right)/2$.

We notice that $p_r(r) \times p_{\mathrm{dir}}(r)  \sim a/\pi$ in the limit $a/r \ll 1$. Hence we obtain that $\GMFPT \sim 2\pi/\left\lbrace 1-\exp(-2\pi/\tau)\right\rbrace$ in this limit. Using the uniform approximation method presented in Sec. \ref{sec:specular_exponential}, we find that the optimal search time reads $\tauopt \sim 2 \pi/\log(\pi/a)$ at leading order in $a \ll 1$. Analysis of simulations  indicate a non-compact scaling for the optimal search time $\GMFPTopt \underset{a \ll 1}{\sim} \pi \xi^{\mathrm{sph}} /(a v),$, where $\xi^{\mathrm{sph}} = 1.0 \pm 0.1$ (see Fig. \ref{Fig6}). We also find that the optimal reorientation time scales as
\begin{align} \label{eq:tauopt_sphere}
\tauopt \underset{a \ll \pi}{\sim} \gamma_{\mathrm{sph}} \frac{\pi}{v\log(\pi/a)},
\end{align}
where $\gamma_{\mathrm{sph}} = 0.5 \pm 0.1$. The uniform approximation provides the appropriate scaling of the optimal reorientation time, however it does not predict the exact value of the proportionality factor $\gamma_{\mathrm{sph}}$. 

We point out that \refn{eq:tauopt_sphere} yields the same scaling as \refn{eq:tauopt}, with however a smaller numerical prefactor.  We suggest that the origin of this discrepancy is related to the following specificity of the spherical geometry: at the South pole ($r \in \left[\pi - a, \pi + a \right]$), all ballistic trajectories eventually cross the target at the North pole (see App. \ref{sec:sphere_geometry}). The optimal reorientation time is shifted to lower values, since shorter reorientation times enhance the probability of tumbles within the South pole region whereby the probability to reach the North pole during the next run is high.

\section{ Extensions and Conclusion } 

To conclude, we have shown that the search time for a target in a bounded domain for  a run and tumble searcher can be minimized by tuning the persistence time. This result is robust and holds for different boundary conditions in 2D and 3D; it was also obtained for deterministic walks and L\'evy walks with finite first moment. This optimal search strategy, which applies to single targets or equivalently regularly spaced targets, therefore notably differs from the  the case of Poisson-distributed targets \cite{Rosa1999}, in which case the straightforward ballistic motion minimizes the search time.  These results therefore highlight the crucial role of the spatial distribution of targets in search problems, and put forward the robustness  of optimal RTWs for single target search problems.

We suggest that our result could also apply to a many-body billiard problem, in which the short-range interactions between freely ballistic particles result in trajectories that can be described as RTWs. The Knudsen regime, in which the mean free path is comparable to the dimension of the confinement \cite{Malek2001,Blanco:2006}, is the analogue of the ballistic limit. In this context, as the mean free path is controlled by the density of particles, our main result suggests that an optimal density could enhance the reaction kinetics of a tracer particle with a fixed target.

%\bibliographystyle{apsrev4-1}
%\bibliography{210514}

\begin{thebibliography}{38}%
\makeatletter
\providecommand \@ifxundefined [1]{%
 \@ifx{#1\undefined}
}%
\providecommand \@ifnum [1]{%
 \ifnum #1\expandafter \@firstoftwo
 \else \expandafter \@secondoftwo
 \fi
}%
\providecommand \@ifx [1]{%
 \ifx #1\expandafter \@firstoftwo
 \else \expandafter \@secondoftwo
 \fi
}%
\providecommand \natexlab [1]{#1}%
\providecommand \enquote  [1]{``#1''}%
\providecommand \bibnamefont  [1]{#1}%
\providecommand \bibfnamefont [1]{#1}%
\providecommand \citenamefont [1]{#1}%
\providecommand \href@noop [0]{\@secondoftwo}%
\providecommand \href [0]{\begingroup \@sanitize@url \@href}%
\providecommand \@href[1]{\@@startlink{#1}\@@href}%
\providecommand \@@href[1]{\endgroup#1\@@endlink}%
\providecommand \@sanitize@url [0]{\catcode `\\12\catcode `\$12\catcode
  `\&12\catcode `\#12\catcode `\^12\catcode `\_12\catcode `\%12\relax}%
\providecommand \@@startlink[1]{}%
\providecommand \@@endlink[0]{}%
\providecommand \url  [0]{\begingroup\@sanitize@url \@url }%
\providecommand \@url [1]{\endgroup\@href {#1}{\urlprefix }}%
\providecommand \urlprefix  [0]{URL }%
\providecommand \Eprint [0]{\href }%
\providecommand \doibase [0]{http://dx.doi.org/}%
\providecommand \selectlanguage [0]{\@gobble}%
\providecommand \bibinfo  [0]{\@secondoftwo}%
\providecommand \bibfield  [0]{\@secondoftwo}%
\providecommand \translation [1]{[#1]}%
\providecommand \BibitemOpen [0]{}%
\providecommand \bibitemStop [0]{}%
\providecommand \bibitemNoStop [0]{.\EOS\space}%
\providecommand \EOS [0]{\spacefactor3000\relax}%
\providecommand \BibitemShut  [1]{\csname bibitem#1\endcsname}%
\let\auto@bib@innerbib\@empty
%</preamble>
\bibitem [{\citenamefont {Einwohner}(1968)}]{Einwohner1968}%
  \BibitemOpen
  \bibfield  {author} {\bibinfo {author} {\bibfnamefont {T.}~\bibnamefont
  {Einwohner}},\ }\href {\doibase 10.1063/1.1670266} {\bibfield  {journal}
  {\bibinfo  {journal} {The Journal of Chemical Physics}\ }\textbf {\bibinfo
  {volume} {49}},\ \bibinfo {pages} {1458} (\bibinfo {year}
  {1968})}\BibitemShut {NoStop}%
\bibitem [{\citenamefont {Knudsen}(1950)}]{Knudsen1950}%
  \BibitemOpen
  \bibfield  {author} {\bibinfo {author} {\bibfnamefont {M.}~\bibnamefont
  {Knudsen}},\ }\href
  {https://books.google.fr/books/about/The{\_}Kinetic{\_}Theory{\_}of{\_}Gases.html?id=OfWfAAAAMAAJ{\&}pgis=1}
  {\emph {\bibinfo {title} {{The Kinetic Theory of Gases: Some Modern
  Aspects}}}}\ (\bibinfo {year} {1950})\ p.~\bibinfo {pages} {64}\BibitemShut
  {NoStop}%
\bibitem [{\citenamefont {Martens}\ \emph {et~al.}(2012)\citenamefont
  {Martens}, \citenamefont {Angelani}, \citenamefont {{Di Leonardo}},\ and\
  \citenamefont {Bocquet}}]{Martens2012}%
  \BibitemOpen
  \bibfield  {author} {\bibinfo {author} {\bibfnamefont {K.}~\bibnamefont
  {Martens}}, \bibinfo {author} {\bibfnamefont {L.}~\bibnamefont {Angelani}},
  \bibinfo {author} {\bibfnamefont {R.}~\bibnamefont {{Di Leonardo}}}, \ and\
  \bibinfo {author} {\bibfnamefont {L.}~\bibnamefont {Bocquet}},\ }\href
  {\doibase 10.1140/epje/i2012-12084-y} {\bibfield  {journal} {\bibinfo
  {journal} {European Physical Journal E}\ }\textbf {\bibinfo {volume} {35}},\
  \bibinfo {pages} {1} (\bibinfo {year} {2012})}\BibitemShut {NoStop}%
\bibitem [{\citenamefont {Moll}\ \emph {et~al.}(2016)\citenamefont {Moll},
  \citenamefont {Kushwaha}, \citenamefont {Nandi}, \citenamefont {Schmidt},\
  and\ \citenamefont {Mackenzie}}]{Moll2016}%
  \BibitemOpen
  \bibfield  {author} {\bibinfo {author} {\bibfnamefont {P.~J.~W.}\
  \bibnamefont {Moll}}, \bibinfo {author} {\bibfnamefont {P.}~\bibnamefont
  {Kushwaha}}, \bibinfo {author} {\bibfnamefont {N.}~\bibnamefont {Nandi}},
  \bibinfo {author} {\bibfnamefont {B.}~\bibnamefont {Schmidt}}, \ and\
  \bibinfo {author} {\bibfnamefont {A.~P.}\ \bibnamefont {Mackenzie}},\ }\href
  {\doibase 10.1126/science.aac8385} {\bibfield  {journal} {\bibinfo  {journal}
  {Science}\ }\textbf {\bibinfo {volume} {351}},\ \bibinfo {pages} {1061}
  (\bibinfo {year} {2016})}\BibitemShut {NoStop}%
\bibitem [{\citenamefont {Zaanen}(2016)}]{Zaanen2016}%
  \BibitemOpen
  \bibfield  {author} {\bibinfo {author} {\bibfnamefont {J.}~\bibnamefont
  {Zaanen}},\ }\href {\doibase 10.1126/science.aaf2487} {\bibfield  {journal}
  {\bibinfo  {journal} {Science (New York, N.Y.)}\ }\textbf {\bibinfo {volume}
  {351}},\ \bibinfo {pages} {1026} (\bibinfo {year} {2016})}\BibitemShut
  {NoStop}%
\bibitem [{\citenamefont {Zoia}\ \emph
  {et~al.}(2011{\natexlab{a}})\citenamefont {Zoia}, \citenamefont {Dumonteil},\
  and\ \citenamefont {Mazzolo}}]{Zoia:2011}%
  \BibitemOpen
  \bibfield  {author} {\bibinfo {author} {\bibfnamefont {A.}~\bibnamefont
  {Zoia}}, \bibinfo {author} {\bibfnamefont {E.}~\bibnamefont {Dumonteil}}, \
  and\ \bibinfo {author} {\bibfnamefont {A.}~\bibnamefont {Mazzolo}},\ }\href
  {http://link.aps.org/doi/10.1103/PhysRevLett.106.220602} {\bibfield
  {journal} {\bibinfo  {journal} {Physical Review Letters}\ }\textbf {\bibinfo
  {volume} {106}} \bibinfo {pages}
  {220602} (\bibinfo {year} {2011}{\natexlab{a}})}\BibitemShut {NoStop}%
\bibitem [{\citenamefont {Zoia}\ \emph
  {et~al.}(2011{\natexlab{b}})\citenamefont {Zoia}, \citenamefont {Dumonteil},\
  and\ \citenamefont {Mazzolo}}]{Zoia:2011a}%
  \BibitemOpen
  \bibfield  {author} {\bibinfo {author} {\bibfnamefont {A.}~\bibnamefont
  {Zoia}}, \bibinfo {author} {\bibfnamefont {E.}~\bibnamefont {Dumonteil}}, \
  and\ \bibinfo {author} {\bibfnamefont {A.}~\bibnamefont {Mazzolo}},\ }\href
  {http://link.aps.org/doi/10.1103/PhysRevE.83.041137} {\bibfield  {journal}
  {\bibinfo  {journal} {Physical Review E}\ }\textbf {\bibinfo {volume} {83}} \bibinfo {pages}
  {041137}
  (\bibinfo {year} {2011}{\natexlab{b}})}\BibitemShut {NoStop}%
\bibitem [{\citenamefont {Romanczuk}\ \emph {et~al.}(2012)\citenamefont
  {Romanczuk}, \citenamefont {Bar}, \citenamefont {Ebeling}, \citenamefont
  {Lindner},\ and\ \citenamefont {Schimansky-Geier}}]{Romanczuk:2012fk}%
  \BibitemOpen
  \bibfield  {author} {\bibinfo {author} {\bibfnamefont {P.}~\bibnamefont
  {Romanczuk}}, \bibinfo {author} {\bibfnamefont {M.}~\bibnamefont {Bar}},
  \bibinfo {author} {\bibfnamefont {W.}~\bibnamefont {Ebeling}}, \bibinfo
  {author} {\bibfnamefont {B.}~\bibnamefont {Lindner}}, \ and\ \bibinfo
  {author} {\bibfnamefont {L.}~\bibnamefont {Schimansky-Geier}},\ }\href
  {\doibase 10.1140/epjst/e2012-01529-y} {\bibfield  {journal} {\bibinfo
  {journal} {EPJE-ST}\ }\textbf {\bibinfo {volume} {202}},\ \bibinfo {pages}
  {1} (\bibinfo {year} {2012})}\BibitemShut {NoStop}%
\bibitem [{\citenamefont {Berg}(2004)}]{Berg2004}%
  \BibitemOpen
  \bibinfo {editor} {\bibfnamefont {H.~C.}\ \bibnamefont {Berg}},\ ed.,\ \href
  {\doibase 10.1007/b97370} {\emph {\bibinfo {title} {PloS one}}},\ \bibinfo
  {series} {Biological and Medical Physics, Biomedical Engineering},
  Vol.~\bibinfo {volume} {7}\ (\bibinfo  {publisher} {Springer New York},\
  \bibinfo {address} {New York, NY},\ \bibinfo {year} {2004})\ p.\ \bibinfo
  {pages} {133}\BibitemShut {NoStop}%
\bibitem [{\citenamefont {J{\"{u}}licher}\ \emph {et~al.}(1997)\citenamefont
  {J{\"{u}}licher}, \citenamefont {Ajdari}, \citenamefont {Prost},\ and\
  \citenamefont {Julicher}}]{Julicher1997}%
  \BibitemOpen
  \bibfield  {author} {\bibinfo {author} {\bibfnamefont {F.}~\bibnamefont
  {J{\"{u}}licher}}, \bibinfo {author} {\bibfnamefont {A.}~\bibnamefont
  {Ajdari}}, \bibinfo {author} {\bibfnamefont {J.}~\bibnamefont {Prost}}, \
  and\ \bibinfo {author} {\bibfnamefont {F.}~\bibnamefont {Julicher}},\ }\href
  {\doibase 10.1103/RevModPhys.69.1269} {\bibfield  {journal} {\bibinfo
  {journal} {Reviews of Modern Physics}\ }\textbf {\bibinfo {volume} {69}},\
  \bibinfo {pages} {1269} (\bibinfo {year} {1997})}\BibitemShut {NoStop}%
\bibitem [{\citenamefont {Heuz{\'{e}}}\ \emph {et~al.}(2013)\citenamefont
  {Heuz{\'{e}}}, \citenamefont {Vargas}, \citenamefont {Chabaud}, \citenamefont
  {{Le Berre}}, \citenamefont {Liu}, \citenamefont {Collin}, \citenamefont
  {Solanes}, \citenamefont {Voituriez}, \citenamefont {Piel},\ and\
  \citenamefont {Lennon-Dum{\'{e}}nil}}]{Heuze:2013uq}%
  \BibitemOpen
  \bibfield  {author} {\bibinfo {author} {\bibfnamefont {M.~L.}\ \bibnamefont
  {Heuz{\'{e}}}}, \bibinfo {author} {\bibfnamefont {P.}~\bibnamefont {Vargas}},
  \bibinfo {author} {\bibfnamefont {M.}~\bibnamefont {Chabaud}}, \bibinfo
  {author} {\bibfnamefont {M.}~\bibnamefont {{Le Berre}}}, \bibinfo {author}
  {\bibfnamefont {Y.-J.}\ \bibnamefont {Liu}}, \bibinfo {author} {\bibfnamefont
  {O.}~\bibnamefont {Collin}}, \bibinfo {author} {\bibfnamefont
  {P.}~\bibnamefont {Solanes}}, \bibinfo {author} {\bibfnamefont
  {R.}~\bibnamefont {Voituriez}}, \bibinfo {author} {\bibfnamefont
  {M.}~\bibnamefont {Piel}}, \ and\ \bibinfo {author} {\bibfnamefont {A.-M.}\
  \bibnamefont {Lennon-Dum{\'{e}}nil}},\ }\href {\doibase 10.1111/imr.12108}
  {\bibfield  {journal} {\bibinfo  {journal} {Immunol Rev}\ }\textbf {\bibinfo
  {volume} {256}},\ \bibinfo {pages} {240} (\bibinfo {year}
  {2013})}\BibitemShut {NoStop}%
\bibitem [{\citenamefont {B{\'{e}}nichou}\ \emph {et~al.}(2011)\citenamefont
  {B{\'{e}}nichou}, \citenamefont {Loverdo}, \citenamefont {Moreau},\ and\
  \citenamefont {Voituriez}}]{Benichou:2011b}%
  \BibitemOpen
  \bibfield  {author} {\bibinfo {author} {\bibfnamefont {O.}~\bibnamefont
  {B{\'{e}}nichou}}, \bibinfo {author} {\bibfnamefont {C.}~\bibnamefont
  {Loverdo}}, \bibinfo {author} {\bibfnamefont {M.}~\bibnamefont {Moreau}}, \
  and\ \bibinfo {author} {\bibfnamefont {R.}~\bibnamefont {Voituriez}},\ }\href
  {http://link.aps.org/doi/10.1103/RevModPhys.83.81} {\bibfield  {journal}
  {\bibinfo  {journal} {Reviews of Modern Physics}\ }\textbf {\bibinfo {volume}
  {83}},\ \bibinfo {pages} {81} (\bibinfo {year} {2011})}\BibitemShut {NoStop}%
\bibitem [{\citenamefont {Shlesinger}(2009)}]{Shlesinger2009}%
  \BibitemOpen
  \bibfield  {author} {\bibinfo {author} {\bibfnamefont {M.~F.}\ \bibnamefont
  {Shlesinger}},\ }\href {\doibase 10.1088/1751-8113/42/43/434001} {\bibfield
  {journal} {\bibinfo  {journal} {Journal of Physics A: Mathematical and
  Theoretical}\ }\textbf {\bibinfo {volume} {42}},\ \bibinfo {pages} {434001}
  (\bibinfo {year} {2009})}\BibitemShut {NoStop}%
  %%
\bibitem [{\citenamefont {Yang}\ \emph {et~al.}(2015)\citenamefont {Yang},
  \citenamefont {Kupka}, \citenamefont {Schuss},\ and\ \citenamefont
  {Holcman}}]{Yang2015}%
  \BibitemOpen
  \bibfield  {author} {\bibinfo {author} {\bibfnamefont {J.}~\bibnamefont
  {Yang}}, \bibinfo {author} {\bibfnamefont {I.}~\bibnamefont {Kupka}},
  \bibinfo {author} {\bibfnamefont {Z.}~\bibnamefont {Schuss}}, \ and\ \bibinfo
  {author} {\bibfnamefont {D.}~\bibnamefont {Holcman}},\ }\href {\doibase
  10.1007/s00285-015-0955-3} {\bibfield  {journal} {\bibinfo  {journal}
  {Journal of Mathematical Biology}\ } \textbf {\bibinfo
  {volume} {73}},\ \bibinfo {pages} {423} (\bibinfo {year} {2016}).}\BibitemShut {NoStop}%
\bibitem [{\citenamefont {Meerson}\ and\ \citenamefont
  {Redner}(2015)}]{Meerson2015a}%
  \BibitemOpen
  \bibfield  {author} {\bibinfo {author} {\bibfnamefont {B.}~\bibnamefont
  {Meerson}}\ and\ \bibinfo {author} {\bibfnamefont {S.}~\bibnamefont
  {Redner}},\ }\href {\doibase 10.1103/PhysRevLett.114.198101} {\bibfield
  {journal} {\bibinfo  {journal} {Physical Review Letters}\ }\textbf {\bibinfo
  {volume} {114}},\ \bibinfo {pages} { 198101} (\bibinfo {year} {2015})}
  \BibitemShut {NoStop}%
  %%
\bibitem [{\citenamefont {Viswanathan}\ \emph {et~al.}(1999)\citenamefont
  {Viswanathan}, \citenamefont {Buldyrev}, \citenamefont {Havlin},
  \citenamefont {da~Luz}, \citenamefont {Raposo},\ and\ \citenamefont
  {Stanley}}]{Rosa1999}%
  \BibitemOpen
  \bibfield  {author} {\bibinfo {author} {\bibfnamefont {G.~M.}\ \bibnamefont
  {Viswanathan}}, \bibinfo {author} {\bibfnamefont {S.~V.}\ \bibnamefont
  {Buldyrev}}, \bibinfo {author} {\bibfnamefont {S.}~\bibnamefont {Havlin}},
  \bibinfo {author} {\bibfnamefont {M.~G.~E.}\ \bibnamefont {da~Luz}}, \bibinfo
  {author} {\bibfnamefont {E.~P.}\ \bibnamefont {Raposo}}, \ and\ \bibinfo
  {author} {\bibfnamefont {H.~E.}\ \bibnamefont {Stanley}},\ }\href
  {http://dx.doi.org/10.1038/44831} {\bibfield  {journal} {\bibinfo  {journal}
  {Nature}\ }\textbf {\bibinfo {volume} {401}},\ \bibinfo {pages} {911}
  (\bibinfo {year} {1999})}\BibitemShut {NoStop}%
\bibitem [{\citenamefont {Edwards}\ \emph {et~al.}(2007)\citenamefont
  {Edwards}, \citenamefont {Phillips}, \citenamefont {Watkins}, \citenamefont
  {Freeman}, \citenamefont {Murphy}, \citenamefont {Afanasyev}, \citenamefont
  {Buldyrev}, \citenamefont {da~Luz}, \citenamefont {Raposo}, \citenamefont
  {Stanley},\ and\ \citenamefont {Viswanathan}}]{Edwards:2007}%
  \BibitemOpen
  \bibfield  {author} {\bibinfo {author} {\bibfnamefont {A.~M.}\ \bibnamefont
  {Edwards}}, \bibinfo {author} {\bibfnamefont {R.~A.}\ \bibnamefont
  {Phillips}}, \bibinfo {author} {\bibfnamefont {N.~W.}\ \bibnamefont
  {Watkins}}, \bibinfo {author} {\bibfnamefont {M.~P.}\ \bibnamefont
  {Freeman}}, \bibinfo {author} {\bibfnamefont {E.~J.}\ \bibnamefont {Murphy}},
  \bibinfo {author} {\bibfnamefont {V.}~\bibnamefont {Afanasyev}}, \bibinfo
  {author} {\bibfnamefont {S.~V.}\ \bibnamefont {Buldyrev}}, \bibinfo {author}
  {\bibfnamefont {M.~G.~E.}\ \bibnamefont {da~Luz}}, \bibinfo {author}
  {\bibfnamefont {E.~P.}\ \bibnamefont {Raposo}}, \bibinfo {author}
  {\bibfnamefont {H.~E.}\ \bibnamefont {Stanley}}, \ and\ \bibinfo {author}
  {\bibfnamefont {G.~M.}\ \bibnamefont {Viswanathan}},\ }\href
  {http://dx.doi.org/10.1038/nature06199} {\bibfield  {journal} {\bibinfo
  {journal} {Nature}\ }\textbf {\bibinfo {volume} {449}},\ \bibinfo {pages}
  {1044} (\bibinfo {year} {2007})}\BibitemShut {NoStop}%
\bibitem [{\citenamefont {Benhamou}(2007)}]{Benhamou:2007fk}%
  \BibitemOpen
  \bibfield  {author} {\bibinfo {author} {\bibfnamefont {S.}~\bibnamefont
  {Benhamou}},\ }\href {\doibase 10.1890/06-1769.1} {\bibfield  {journal}
  {\bibinfo  {journal} {Ecology}\ }\textbf {\bibinfo {volume} {88}},\ \bibinfo
  {pages} {1962} (\bibinfo {year} {2007})}\BibitemShut {NoStop}%
\bibitem [{\citenamefont {James}\ \emph {et~al.}(2011)\citenamefont {James},
  \citenamefont {Plank},\ and\ \citenamefont {Edwards}}]{James:2011}%
  \BibitemOpen
  \bibfield  {author} {\bibinfo {author} {\bibfnamefont {A.}~\bibnamefont
  {James}}, \bibinfo {author} {\bibfnamefont {M.~J.}\ \bibnamefont {Plank}}, \
  and\ \bibinfo {author} {\bibfnamefont {A.~M.}\ \bibnamefont {Edwards}},\
  }\href
  {http://rsif.royalsocietypublishing.org/content/early/2011/05/25/rsif.2011.0200.abstract}
  {\bibfield  {journal} {\bibinfo  {journal} {Journal of The Royal Society
  Interface}\ } \textbf {\bibinfo {volume} {8}},\ \bibinfo
  {pages} {1233} (\bibinfo {year} {2011})}\BibitemShut {NoStop}%
\bibitem [{\citenamefont {Palyulin}\ \emph {et~al.}(2014)\citenamefont
  {Palyulin}, \citenamefont {Chechkin},\ and\ \citenamefont
  {Metzler}}]{Palyulin2014}%
  \BibitemOpen
  \bibfield  {author} {\bibinfo {author} {\bibfnamefont {V.~V.}\ \bibnamefont
  {Palyulin}}, \bibinfo {author} {\bibfnamefont {A.~V.}\ \bibnamefont
  {Chechkin}}, \ and\ \bibinfo {author} {\bibfnamefont {R.}~\bibnamefont
  {Metzler}},\ }\href {\doibase 10.1073/pnas.1320424111} {\bibfield  {journal}
  {\bibinfo  {journal} {Proceedings of the National Academy of Sciences of the
  United States of America}\ }\textbf {\bibinfo {volume} {111}},\ \bibinfo
  {pages} {2931} (\bibinfo {year} {2014})}\BibitemShut {NoStop}%
\bibitem [{\citenamefont {Condamin}\ \emph {et~al.}(2007)\citenamefont
  {Condamin}, \citenamefont {B{\'{e}}nichou}, \citenamefont {Tejedor},
  \citenamefont {Voituriez},\ and\ \citenamefont {Klafter}}]{Condamin:2007zl}%
  \BibitemOpen
  \bibfield  {author} {\bibinfo {author} {\bibfnamefont {S.}~\bibnamefont
  {Condamin}}, \bibinfo {author} {\bibfnamefont {O.}~\bibnamefont
  {B{\'{e}}nichou}}, \bibinfo {author} {\bibfnamefont {V.}~\bibnamefont
  {Tejedor}}, \bibinfo {author} {\bibfnamefont {R.}~\bibnamefont {Voituriez}},
  \ and\ \bibinfo {author} {\bibfnamefont {J.}~\bibnamefont {Klafter}},\ }\href
  {\doibase 10.1038/nature06201} {\bibfield  {journal} {\bibinfo  {journal}
  {Nature}\ }\textbf {\bibinfo {volume} {450}},\ \bibinfo {pages} {77}
  (\bibinfo {year} {2007})}\BibitemShut {NoStop}%
\bibitem [{\citenamefont {B{\'{e}}nichou}\ \emph {et~al.}(2010)\citenamefont
  {B{\'{e}}nichou}, \citenamefont {Chevalier}, \citenamefont {Klafter},
  \citenamefont {Meyer},\ and\ \citenamefont {Voituriez}}]{BenichouO.:2010a}%
  \BibitemOpen
  \bibfield  {author} {\bibinfo {author} {\bibfnamefont {O.}~\bibnamefont
  {B{\'{e}}nichou}}, \bibinfo {author} {\bibfnamefont {C.}~\bibnamefont
  {Chevalier}}, \bibinfo {author} {\bibfnamefont {J.}~\bibnamefont {Klafter}},
  \bibinfo {author} {\bibfnamefont {B.}~\bibnamefont {Meyer}}, \ and\ \bibinfo
  {author} {\bibfnamefont {R.}~\bibnamefont {Voituriez}},\ }\href
  {https://acces-distant.upmc.fr:443/http/dx.doi.org/10.1038/nchem.622}
  {\bibfield  {journal} {\bibinfo  {journal} {Nat Chem}\ }\textbf {\bibinfo
  {volume} {2}},\ \bibinfo {pages} {472} (\bibinfo {year} {2010})}\BibitemShut
  {NoStop}%
\bibitem [{\citenamefont {B{\'{e}}nichou}\ and\ \citenamefont
  {Voituriez}(2014)}]{Benichou2014}%
  \BibitemOpen
  \bibfield  {author} {\bibinfo {author} {\bibfnamefont {O.}~\bibnamefont
  {B{\'{e}}nichou}}\ and\ \bibinfo {author} {\bibfnamefont {R.}~\bibnamefont
  {Voituriez}},\ }\href {\doibase 10.1016/j.physrep.2014.02.003} {\bibfield
  {journal} {\bibinfo  {journal} {Physics Reports}\ }\textbf {\bibinfo {volume}
  {539}},\ \bibinfo {pages} {225} (\bibinfo {year} {2014})}\BibitemShut
  {NoStop}%
\bibitem [{\citenamefont {Gu{\'{e}}rin}\ \emph {et~al.}(2012)\citenamefont
  {Gu{\'{e}}rin}, \citenamefont {B{\'{e}}nichou},\ and\ \citenamefont
  {Voituriez}}]{Guerin2012}%
  \BibitemOpen
  \bibfield  {author} {\bibinfo {author} {\bibfnamefont {T.}~\bibnamefont
  {Gu{\'{e}}rin}}, \bibinfo {author} {\bibfnamefont {O.}~\bibnamefont
  {B{\'{e}}nichou}}, \ and\ \bibinfo {author} {\bibfnamefont {R.}~\bibnamefont
  {Voituriez}},\ }\href {\doibase 10.1038/nchem.1378} {\bibfield  {journal}
  {\bibinfo  {journal} {Nature Chemistry}\ }\textbf {\bibinfo {volume} {4}},\
  \bibinfo {pages} {568} (\bibinfo {year} {2012})}\BibitemShut {NoStop}%
 %%
 \bibitem [{\citenamefont {Singer}\ \emph {et~al.}(2006)\citenamefont {Singer},
  \citenamefont {Schuss}, \citenamefont {Holcman},\ and\ \citenamefont
  {Eisenberg}}]{Singer:2006a}%
  \BibitemOpen
  \bibfield  {author} {\bibinfo {author} {\bibfnamefont {A.}~\bibnamefont
  {Singer}}, \bibinfo {author} {\bibfnamefont {Z.}~\bibnamefont {Schuss}},
  \bibinfo {author} {\bibfnamefont {D.}~\bibnamefont {Holcman}}, \ and\
  \bibinfo {author} {\bibfnamefont {R.}~\bibnamefont {Eisenberg}},\ }\href
  {http://dx.doi.org/10.1007/s10955-005-8026-6} {\bibfield  {journal} {\bibinfo
   {journal} {Journal of Statistical Physics}\ }\textbf {\bibinfo {volume}
  {122}},\ \bibinfo {pages} {437} (\bibinfo {year} {2006})}\BibitemShut
  {NoStop}%
%%
\bibitem [{\citenamefont {Singer}\ \emph {et~al.}(2006)\citenamefont {Singer},
  \citenamefont {Schuss},\ and\ \citenamefont {Holcman}}]{Singer2006}%
  \BibitemOpen
  \bibfield  {author} {\bibinfo {author} {\bibfnamefont {A.}~\bibnamefont
  {Singer}}, \bibinfo {author} {\bibfnamefont {Z.}~\bibnamefont {Schuss}}, \
  and\ \bibinfo {author} {\bibfnamefont {D.}~\bibnamefont {Holcman}},\ }\href
  {\doibase DOI 10.1007/s10955-005-8027-5} {\bibfield  {journal} {\bibinfo
  {journal} {Journal of Statistical Physics}\ }\textbf {\bibinfo {volume}
  {122}},\ \bibinfo {pages} {465} (\bibinfo {year} {2006})}\BibitemShut
  {NoStop}%
\bibitem [{\citenamefont {Schuss}\ \emph {et~al.}(2007)\citenamefont {Schuss},
  \citenamefont {Singer},\ and\ \citenamefont {Holcman}}]{Schuss2007}%
  \BibitemOpen
  \bibfield  {author} {\bibinfo {author} {\bibfnamefont {Z.}~\bibnamefont
  {Schuss}}, \bibinfo {author} {\bibfnamefont {A.}~\bibnamefont {Singer}}, \
  and\ \bibinfo {author} {\bibfnamefont {D.}~\bibnamefont {Holcman}},\ }\href
  {\doibase 10.1073/pnas.0706599104} {\bibfield  {journal} {\bibinfo  {journal}
  {Proceedings of the National Academy of Sciences of the United States of
  America}\ }\textbf {\bibinfo {volume} {104}},\ \bibinfo {pages} {16098}
  (\bibinfo {year} {2007})}\BibitemShut {NoStop}%  
%%  
  \bibitem [{\citenamefont {Rupprecht}\ \emph {et~al.}(2014)\citenamefont
  {Rupprecht}, \citenamefont {B{\'{e}}nichou}, \citenamefont {Grebenkov},\ and\
  \citenamefont {Voituriez}}]{Rupprecht2014}%
  \BibitemOpen
  \bibfield  {author} {\bibinfo {author} {\bibfnamefont {J.-F.}\ \bibnamefont
  {Rupprecht}}, \bibinfo {author} {\bibfnamefont {O.}~\bibnamefont
  {B{\'{e}}nichou}}, \bibinfo {author} {\bibfnamefont {D.~S.}\ \bibnamefont
  {Grebenkov}}, \ and\ \bibinfo {author} {\bibfnamefont {R.}~\bibnamefont
  {Voituriez}},\ }\href {\doibase 10.1007/s10955-014-1116-6} {\textbf {\bibinfo {volume} {158}},\
  \bibinfo {pages} {192} \bibfield
  {journal} {\bibinfo  {journal} {Journal of Statistical Physics}\ } (\bibinfo
  {year} {2014})}\BibitemShut {NoStop}%
%%  
\bibitem [{\citenamefont {Isaacson}\ and\ \citenamefont
  {Newby}(2013)}]{Isaacson2013a}%
  \BibitemOpen
  \bibfield  {author} {\bibinfo {author} {\bibfnamefont {S.~A.}\ \bibnamefont
  {Isaacson}}\ and\ \bibinfo {author} {\bibfnamefont {J.}~\bibnamefont
  {Newby}},\ }\href {\doibase 10.1103/PhysRevE.88.012820} {\bibfield  {journal}
  {\bibinfo  {journal} {Physical Review E}\ }\textbf {\bibinfo {volume} {88}},\
  \bibinfo {pages} {012820} (\bibinfo {year} {2013})}\BibitemShut {NoStop}%
%%
\bibitem [{\citenamefont {Gilbert}\ \emph {et~al.}(2011)\citenamefont
  {Gilbert}, \citenamefont {Nguyen},\ and\ \citenamefont
  {Sanders}}]{Gilbert:2011}%
  \BibitemOpen
  \bibfield  {author} {\bibinfo {author} {\bibfnamefont {T.}~\bibnamefont
  {Gilbert}}, \bibinfo {author} {\bibfnamefont {H.~C.}\ \bibnamefont {Nguyen}},
  \ and\ \bibinfo {author} {\bibfnamefont {D.~P.}\ \bibnamefont {Sanders}},\
  }\href {http://stacks.iop.org/1751-8121/44/i=6/a=065001} {\bibfield
  {journal} {\bibinfo  {journal} {Journal of Physics A: Mathematical and
  Theoretical}\ }\textbf {\bibinfo {volume} {44}} (\bibinfo {year}
  {2011})}\BibitemShut {NoStop}%
\bibitem [{\citenamefont {Blanco}\ and\ \citenamefont
  {Fournier}(2003)}]{Blanco:2003a}%
  \BibitemOpen
  \bibfield  {author} {\bibinfo {author} {\bibfnamefont {S.}~\bibnamefont
  {Blanco}}\ and\ \bibinfo {author} {\bibfnamefont {R.}~\bibnamefont
  {Fournier}},\ }\href@noop {} {\bibfield  {journal} {\bibinfo  {journal} {EPL
  (Europhysics Letters)}\ }\textbf {\bibinfo {volume} {61}},\ \bibinfo {pages}
  {168} (\bibinfo {year} {2003})}\BibitemShut {NoStop}%
\bibitem [{\citenamefont {B{\'{e}}nichou}\ \emph {et~al.}(2005)\citenamefont
  {B{\'{e}}nichou}, \citenamefont {Coppey}, \citenamefont {Moreau},
  \citenamefont {Suet},\ and\ \citenamefont {Voituriez}}]{Benichou:2005a}%
  \BibitemOpen
  \bibfield  {author} {\bibinfo {author} {\bibfnamefont {O.}~\bibnamefont
  {B{\'{e}}nichou}}, \bibinfo {author} {\bibfnamefont {M.}~\bibnamefont
  {Coppey}}, \bibinfo {author} {\bibfnamefont {M.}~\bibnamefont {Moreau}},
  \bibinfo {author} {\bibfnamefont {P.~H.}\ \bibnamefont {Suet}}, \ and\
  \bibinfo {author} {\bibfnamefont {R.}~\bibnamefont {Voituriez}},\ }\href@noop
  {} {\bibfield  {journal} {\bibinfo  {journal} {EPL (Europhysics Letters)}\
  }\textbf {\bibinfo {volume} {70}},\ \bibinfo {pages} {42} (\bibinfo {year}
  {2005})}\BibitemShut {NoStop}%
\bibitem [{\citenamefont {B{\'{e}}nichou}\ \emph {et~al.}(2006)\citenamefont
  {B{\'{e}}nichou}, \citenamefont {Loverdo}, \citenamefont {Moreau},\ and\
  \citenamefont {Voituriez}}]{Benichou2006}%
  \BibitemOpen
  \bibfield  {author} {\bibinfo {author} {\bibfnamefont {O.}~\bibnamefont
  {B{\'{e}}nichou}}, \bibinfo {author} {\bibfnamefont {C.}~\bibnamefont
  {Loverdo}}, \bibinfo {author} {\bibfnamefont {M.}~\bibnamefont {Moreau}}, \
  and\ \bibinfo {author} {\bibfnamefont {R.}~\bibnamefont {Voituriez}},\ }\href
  {\doibase 10.1103/PhysRevE.74.020102} {\bibfield  {journal} {\bibinfo
  {journal} {Physical Review E}\ }\textbf {\bibinfo {volume} {74}},\ \bibinfo
  {pages} {020102} (\bibinfo {year} {2006})}\BibitemShut {NoStop}%
\bibitem [{\citenamefont {Benichou}\ \emph {et~al.}(2007)\citenamefont
  {Benichou}, \citenamefont {Loverdo}, \citenamefont {Moreau},\ and\
  \citenamefont {Voituriez}}]{Benichou:2007}%
  \BibitemOpen
  \bibfield  {author} {\bibinfo {author} {\bibfnamefont {O.}~\bibnamefont
  {Benichou}}, \bibinfo {author} {\bibfnamefont {C.}~\bibnamefont {Loverdo}},
  \bibinfo {author} {\bibfnamefont {M.}~\bibnamefont {Moreau}}, \ and\ \bibinfo
  {author} {\bibfnamefont {R.}~\bibnamefont {Voituriez}},\ }\href@noop {}
  {\bibfield  {journal} {\bibinfo  {journal} {Journal Of Physics-Condensed
  Matter}\ }\textbf {\bibinfo {volume} {19}},\ \bibinfo {pages} {65141}
  (\bibinfo {year} {2007})}\BibitemShut {NoStop}%
\bibitem [{\citenamefont {Tejedor}\ \emph {et~al.}(2012)\citenamefont
  {Tejedor}, \citenamefont {Voituriez},\ and\ \citenamefont
  {B{\'{e}}nichou}}]{Tejedor:2012ly}%
  \BibitemOpen
  \bibfield  {author} {\bibinfo {author} {\bibfnamefont {V.}~\bibnamefont
  {Tejedor}}, \bibinfo {author} {\bibfnamefont {R.}~\bibnamefont {Voituriez}},
  \ and\ \bibinfo {author} {\bibfnamefont {O.}~\bibnamefont {B{\'{e}}nichou}},\
  }\href {http://link.aps.org/doi/10.1103/PhysRevLett.108.088103} {\bibfield
  {journal} {\bibinfo  {journal} {Physical Review Letters}\ }\textbf {\bibinfo
  {volume} {108}},\ \bibinfo {pages} {088103} (\bibinfo {year}
  {2012})}\BibitemShut {NoStop}%
      %%%%
   \BibitemOpen
\bibitem [{\citenamefont {Campos}\ \emph {et~al.}(2015)\citenamefont {Campos},
  \citenamefont {Abad}, \citenamefont {Mendez}, \citenamefont {Yuste},\ and\
  \citenamefont {Lindenberg}}]{Campos2015a}%
  \BibitemOpen
  \bibfield  {author} {\bibinfo {author} {\bibfnamefont {D.}~\bibnamefont
  {Campos}}, \bibinfo {author} {\bibfnamefont {E.}~\bibnamefont {Abad}},
  \bibinfo {author} {\bibfnamefont {V.}~\bibnamefont {Mendez}}, \bibinfo
  {author} {\bibfnamefont {S.~B.}\ \bibnamefont {Yuste}}, \ and\ \bibinfo
  {author} {\bibfnamefont {K.}~\bibnamefont {Lindenberg}},\ }\href {\doibase
  10.1103/PhysRevE.91.052115} {\bibfield  {journal} {\bibinfo  {journal}
  {Physical Review E - Statistical, Nonlinear, and Soft Matter Physics}\
  }\textbf {\bibinfo {volume} {91}},\ \bibinfo {pages} {1} (\bibinfo {year}
  {2015})}
  \BibitemShut {NoStop}%
  %%
  \bibitem [{\citenamefont {Romanczuk}\ and\ \citenamefont
  {Salbreux}(2015)}]{Romanczuk2015}%
  \BibitemOpen
  \bibfield  {author} {\bibinfo {author} {\bibfnamefont {P.}~\bibnamefont
  {Romanczuk}}\ and\ \bibinfo {author} {\bibfnamefont {G.}~\bibnamefont
  {Salbreux}}}
  \href {\doibase 10.1103/PhysRevE.91.042720} {\bibinfo  {journal} {Physical Review E} \ \textbf
  {\bibinfo {volume} { 91}},\ \bibinfo {pages} { 042720} (\bibinfo {year}
  {2015})}\BibitemShut {NoStop}%
  \bibitem [{\citenamefont {Wang}\ \emph {et~al.}(2016)\citenamefont {Wang},
  \citenamefont {Chen}, \citenamefont {Yu},\ and\ \citenamefont
  {Luo}}]{Wang2016}%
  \BibitemOpen
  \bibfield  {author} {\bibinfo {author} {\bibfnamefont {J.}~\bibnamefont
  {Wang}}, \bibinfo {author} {\bibfnamefont {Y.}~\bibnamefont {Chen}}, \bibinfo
  {author} {\bibfnamefont {W.}~\bibnamefont {Yu}}, \ and\ \bibinfo {author}
  {\bibfnamefont {K.}~\bibnamefont {Luo}},\ }\href {\doibase 10.1063/1.4952423}
  {\bibfield  {journal} {\bibinfo  {journal} {The Journal of Chemical Physics}\
  }\textbf {\bibinfo {volume} {144}},\ \bibinfo {pages} {204702} (\bibinfo
  {year} {2016})}\BibitemShut {NoStop}%
  %%
\bibitem [{\citenamefont {Gardiner}(2009)}]{Gardiner2009}%
  \BibitemOpen
  \bibfield  {author} {\bibinfo {author} {\bibfnamefont {C.}~\bibnamefont
  {Gardiner}},\ }\href@noop {} {\emph {\bibinfo {title} {{Stochastic Methods: A
  Handbook for the Natural and Social Sciences}}}}\ (\bibinfo  {publisher}
  {Springer},\ \bibinfo {year} {2009})\ p.\ \bibinfo {pages} {447}\BibitemShut
  {NoStop}%
\bibitem [{\citenamefont {Weiss}(1984)}]{Weiss1984}%
  \BibitemOpen
  \bibfield  {author} {\bibinfo {author} {\bibfnamefont {G.~H.}\ \bibnamefont
  {Weiss}},\ }\href {\doibase 10.1007/BF01011837} {\bibfield  {journal}
  {\bibinfo  {journal} {Journal of Statistical Physics}\ }\textbf {\bibinfo
  {volume} {37}},\ \bibinfo {pages} {325} (\bibinfo {year} {1984})}\BibitemShut
  {NoStop}%
\bibitem [{\citenamefont {Redner}(2001)}]{Redner:2001a}%
  \BibitemOpen
  \bibfield  {author} {\bibinfo {author} {\bibfnamefont {S.}~\bibnamefont
  {Redner}},\ }\href@noop {} {\emph {\bibinfo {title} {{A guide to First-
  Passage Processes}}}}\ (\bibinfo  {publisher} {Cambridge University Press,
  Cambridge, England},\ \bibinfo {year} {2001})\BibitemShut {NoStop}%
\bibitem [{\citenamefont {Feres}\ and\ \citenamefont {Yablonsky}()}]{Feres}%
  \BibitemOpen
  \bibfield  {author} {\bibinfo {author} {\bibfnamefont {R.}~\bibnamefont
  {Feres}}\ and\ \bibinfo {author} {\bibfnamefont {G.}~\bibnamefont
  {Yablonsky}},\ }\href {\doibase 10.1016/j.ces.2004.01.016 } {\bibfield  {journal}
  {\bibinfo  {journal} {Chemical Engineering Science}\ }\textbf {\bibinfo
  {volume} {59}},\ \bibinfo {pages} {1541} (\bibinfo {year} {2004})}\BibitemShut
  {NoStop}%
\bibitem [{\citenamefont {Tailleur}\ and\ \citenamefont
  {Cates}(2008)}]{Tailleur2008}%
  \BibitemOpen
  \bibfield  {author} {\bibinfo {author} {\bibfnamefont {J.}~\bibnamefont
  {Tailleur}}\ and\ \bibinfo {author} {\bibfnamefont {M. E.}~\bibnamefont
  {Cates}},\ }\href {\doibase 10.1103/PhysRevLett.100.218103} {\bibfield
  {journal} {\bibinfo  {journal} {Physical Review Letters}\ }\textbf {\bibinfo
  {volume} {100}},\ \bibinfo {pages} {218103} (\bibinfo {year}
  {2008})}\BibitemShut {NoStop}%
\bibitem [{\citenamefont {Ghosh}\ \emph {et~al.}(2013)\citenamefont {Ghosh},
  \citenamefont {Misko}, \citenamefont {Marchesoni},\ and\ \citenamefont
  {Nori}}]{Ghosh2013}%
  \BibitemOpen
  \bibfield  {author} {\bibinfo {author} {\bibfnamefont {P.~K.}\ \bibnamefont
  {Ghosh}}, \bibinfo {author} {\bibfnamefont {V.~R.}\ \bibnamefont {Misko}},
  \bibinfo {author} {\bibfnamefont {F.}~\bibnamefont {Marchesoni}}, \ and\
  \bibinfo {author} {\bibfnamefont {F.}~\bibnamefont {Nori}},\ }\href {\doibase
  10.1103/PhysRevLett.110.268301} {\bibfield  {journal} {\bibinfo  {journal}
  {Physical Review Letters}\ }\textbf {\bibinfo {volume} {110}},\ \bibinfo
  {pages} {268301} (\bibinfo {year} {2013})},
  \BibitemShut {NoStop}%  
\bibitem [{\citenamefont {Malek}\ and\ \citenamefont
  {Coppens}(2001)}]{Malek2001}%
  \BibitemOpen
  \bibfield  {author} {\bibinfo {author} {\bibfnamefont {K.}~\bibnamefont
  {Malek}}\ and\ \bibinfo {author} {\bibfnamefont {M.~O.}\ \bibnamefont
  {Coppens}},\ }\href {\doibase 10.1103/PhysRevLett.87.125505} {\bibfield
  {journal} {\bibinfo  {journal} {Physical review letters}\ }\textbf {\bibinfo
  {volume} {87}},\ \bibinfo {pages} {125505} (\bibinfo {year}
  {2001})}\BibitemShut {NoStop}%
\bibitem [{\citenamefont {Blanco}\ and\ \citenamefont
  {Fournier}(2006)}]{Blanco:2006}%
  \BibitemOpen
  \bibfield  {author} {\bibinfo {author} {\bibfnamefont {S.}~\bibnamefont
  {Blanco}}\ and\ \bibinfo {author} {\bibfnamefont {R.}~\bibnamefont
  {Fournier}},\ }\href {http://link.aps.org/abstract/PRL/v97/e230604}
  {\bibfield  {journal} {\bibinfo  {journal} {Physical Review Letters}\
  }\textbf {\bibinfo {volume} {97}},\ \bibinfo {pages} {230604} (\bibinfo
  {year} {2006})}\BibitemShut {NoStop}%
\bibitem [{\citenamefont {Harris}\ and\ \citenamefont
  {Stocker}(1999)}]{Harris1999}%
  \BibitemOpen
  \bibfield  {author} {\bibinfo {author} {\bibfnamefont {J.~W.}\ \bibnamefont
  {Harris}}\ and\ \bibinfo {author} {\bibfnamefont {H.}~\bibnamefont
  {Stocker}},\ }\href {\doibase 10.1016/S0898-1221(99)90385-1} {\bibfield
  {journal} {\bibinfo  {journal} {Computers {\&} Mathematics with
  Applications}\ }\textbf {\bibinfo {volume} {37}},\ \bibinfo {pages} {133}
  (\bibinfo {year} {1999})}\BibitemShut {NoStop}%
\end{thebibliography}


\appendix 

\section{Analytical results for a regular run-and-tumble walk} \label{app:regular}
Here, we consider that reorientations occur at regular time intervals $\tau$, i.e. the distribution of reorientation times reads $\pi(t) = \delta(t- \tau)$, where $\delta$ is the Dirac function. We recall that in the presence of a scattering boundary condition at $r=b$, the search time exibits a sharp minimum at $\tau = b/v$. We obtain an analytical description for the minima at $\tau =b/v$ in 2D.  \\

We start by considering that the searcher is started at the scattering boundary $r=b$. After a time $\tau$, the position of the particle is either $\tilde{r}(\theta) = \sqrt{b^2 + v^2 \tau^2 + 2 b v \tau \cos(\theta)} \in \left]a, b\right[$, $\tilde{r}(\theta) =a$ or $\tilde{r}(\theta) =b$. These three situations may occur, depending on the value of the ratio $(v \tau)/b$.

\begin{figure}[t!]
  \includegraphics[width=9cm]{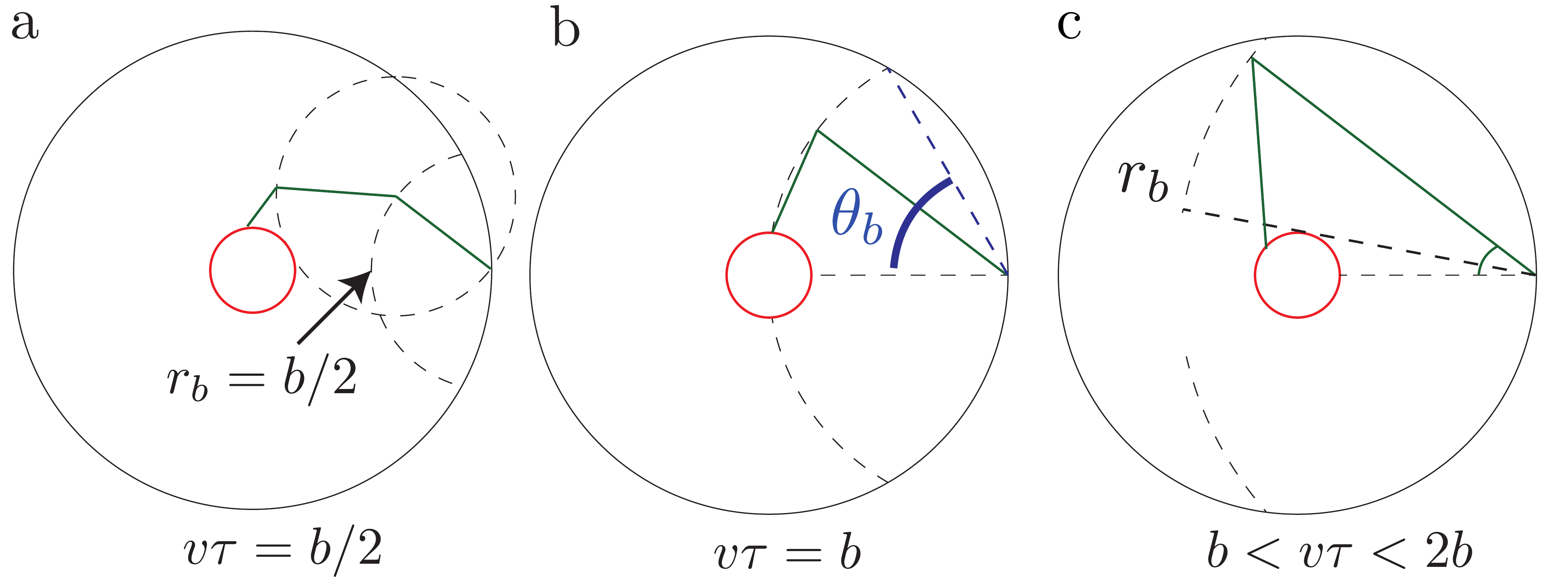}
  \caption{Sketches of the search process. The choice $\tau = b/(n v)$ maximizes the probability to reach the target after the $n$--th reorientation event: (a) $n=2$ (b) $n = 1$, (c) the search time increases when $\tau > b/v$.}
\label{fig:Dirac}
\end{figure}

\paragraph*{First case -- $b < v \tau < 2b$} Here, we distinguish three sub-situations, according to the value of the initial direction $\theta \in \left[0, \pi/2\right]$ (see Fig. \ref{fig:Dirac}). 

If $\pi - \theta_a < \theta < \pi$, where $\theta_a  = \arcsin\left(a/b\right)$, the searcher hits the target in a time $b/v$ (up to a minor term that scales as $a/v$). 

If $\pi - \theta_b < \theta < \pi - \theta_a$, where $\theta_b = \arccos\left[\tau/(2b)\right]$, the searcher is reoriented at the position  $\tilde{r}(\theta) \in \left]a, b\right[$. The final radius of a searcher whose initial direction is $\theta = \theta_a$ is denoted $r_b \equiv \tilde{r}(\theta_a)$ and reads:
\begin{align}
r_b = \sqrt{b^2 - 2\sqrt{b^2 - a^2} v \tau + v^2\tau^2}.
\end{align}
The distribution of final position $\tilde{r}$ after a single run is a non-uniformly distributed random variable, with a density probability equal to the Jacobian:
\begin{align} \label{eq:jacobian}
\nm{\pd{\theta}{\tilde{r} }} = \frac{ 2 \tau \, \tilde{r} }{b \tau  \sqrt{4 b^2 \tau ^2-\left(b^2-\tilde{r} ^2+\tau ^2\right)^2}}.
\end{align}
Notice that the Jacobian defined in \refn{eq:jacobian} diverges for  $b - \tau$ (i.e. $\theta = 0$), which is expected since a small variation of $\tilde{r}$ around $b - \tau$ corresponds to an abrupt change in the value of $\theta$. 

Finally, if $\pi/2 < \theta < \pi - \theta_b$, the searcher takes a time $t_b(\theta) = -2 b \cos(\theta)/v$ to reach its final position at $\tilde{r} = b$. In 2D, the mean return time to the boundary $r=b$, averaged over the angles $\theta \in \left[\theta_b, \pi/2 \right]$, reads:
\begin{align} \label{eq:averagedreturn}
\int^{\pi - \theta_b}_{\pi/2} \! \frac{2 \mathrm{d} \theta}{\pi} \, t_b(\theta)= \frac{2 b \left(2-\sqrt{4-\frac{v^2\tau ^2}{b^2}}\right)}{\pi  v}.
\end{align}

\paragraph*{Second case -- $v \tau < (b-a)$} Here, the final position of the particle $\tilde{r}(\theta)$ takes all values between $b$ and its minimal value $r_b \equiv b - v\tau$. 

\paragraph*{Third case -- $2 b < v \tau$} The search time is a constant of $\tau$ and is strictly equal to the expression of \refn{eq:taularge_diffusive}. Within a single run from the boundary, the target is either hit or is missed and the final position is $\tilde{r}=b$.  We set $\theta_b = \theta_a$ if $2 b < v \tau$.

Let us detail the situation in the case $b < v \tau < 2b$. The renewal equation method (as presented in \cite{Redner:2001a}) leads to the following integral equation on the search time:
\begin{align} \label{eq:renewdirac}
&\left<t(b)\right> = \frac{2 a}{\pi b} \frac{b}{v}  + \int^{b}_{r_b}  \frac{2 \mathrm{d}\tilde{r}}{\pi}  \nm{\pd{\theta}{\tilde{r} }} \left\lbrace \tau + \left<t(\tilde{r})\right> \right\rbrace \nonumber \\
&+ \left[1-\frac{2 \theta_b}{\pi}\right]  \left[  \frac{2 b}{\pi v} \left(2-\sqrt{4-\frac{v^2\tau ^2}{b^2}}\right) \, + \,  \left<t(b)\right> \right],
\end{align}
where (i) the first term in the right hand side of \refn{eq:renewdirac} corresponds to the product of the probability to reach the target in a single run by the characteristic time of the run, 
(ii) the second term corresponds to the reorientation events at a random position $\tilde{r} \in \left[r_b, b\right]$, and (iii) the last term corresponds to the product of the probability to reach the boundary at $\tilde{r} = b$ in a single run by the sum of (a) the average time \refn{eq:averagedreturn} to reach the boundary at $\tilde{r} = b$ and (b) of the averaged search time from $\tilde{r} = b$.

To simplify the expression of \refn{eq:renewdirac}, we make the assumption that
\begin{align} \label{eq:renewlocal}
\left<t(\tilde{r})\right> = \frac{1}{\pi} \frac{a}{\tilde{r}} \frac{\tilde{r}}{v} + \left\lbrace \tau + \left<t(b)\right> \right\rbrace \left[1- \frac{a}{\pi\tilde{r}} \right],
\end{align}
which is justified as (i) the first term on the right hand side of \refn{eq:renewlocal} corresponds to a success to encounter the target after the reorientation event, and (ii) the second term corresponds to a failure to encounter the target after the reorientation event. With the approximation of \refn{eq:renewlocal}, the \refn{eq:renewdirac} becomes:
\begin{align} \label{eq:firstorder_dirac}
\left<t(b)\right> = \frac{\frac{2 a}{b \pi} \frac{b}{v}  + \int^{b}_{r_b}  \frac{2 \mathrm{d}\tilde{r}}{\pi}  \nm{\pd{\theta}{\tilde{r} }} \left( 2\tau + a/v \right)
+ \left[1-\frac{2 \theta_b}{\pi}\right]   t_d}{1 - \int^{b}_{r_b} \! \mathrm{d}\tilde{r}  \nm{\pd{\theta}{\tilde{r} }} \left[1- \frac{a}{\pi \tilde{r}} \right] - \left[1-\frac{2 \theta_b}{\pi}\right] },
\end{align}
where
\begin{align*}
t_d = \left[ \frac{2 b}{\pi v} \left(2-\sqrt{4-\frac{v^2\tau ^2}{b^2}}\right)\right].
\end{align*}
The expression \refn{eq:firstorder_dirac} provides a satisfactory description of the minimum at $\tau = b/v$ (see Fig. \ref{fig:Dirac}. b.). The arguments presented here can be extended to the 3D geometry.

\section{Sphere geometry} \label{sec:sphere_geometry}

We consider a ballistic particle starts moving from its initial position at the zenith angle $r_0$ (on the surface of a sphere of radius $1$), where $\theta_0$ refers to the angle between the velocity vector and the position vector (see Fig. \ref{Fig6}). The equation of motion of this ballistic particle reads \cite{Harris1999}
\begin{align} \label{eq:motionsphere}
\cos r(t) = \cos r_0 \cos t +  \sin r_0 \sin t \cos \theta_0, 
\end{align}
where we have set $v=1$. To determine whether the particle will cross the target, we solve \refn{eq:motionsphere} in terms of $t_a$ with the final condition $r(t_a) = a$ and we find:
\begin{align} \label{eq:ta_si} 
t_a =  \frac{\cos (r) \cos (r_{0})-\sqrt{\sin ^2(r_0) \cos ^2(\theta_{0}) \Delta}}{\sin ^2(r_0) \cos ^2(\theta_{0})+\cos ^2(r_0)},
\end{align}
where
\begin{align*}
\Delta = \cos ^2(r_0) -\cos ^2(r) +\sin ^2(r_0) \cos ^2(\theta_{0}).
\end{align*}
The condition $\Delta >0$ defines the critical angle
\begin{align} \label{eq:si_thetaa}
\theta_a = \arccos\left(\frac{\sqrt{\cos^2 a - \cos^2 r_0}}{\sin r_0}\right).
\end{align}
Hence the trajectory crosses the target if the initial angle $\theta_0$ lies either in the interval $\left[0, \theta_{a} \right]$ or in $\left[\pi-\theta_a, \pi\right]$. Notice that for tumbles occurring in the South pole region $r \in \left[\pi - a, \pi + a \right]$, we have $\theta_a= \pi$, as expected. Indeed, in this region, all ballistic trajectories will eventually cross the target at the North pole.

Along a trajectory between $r$ and $r'$ of length $vt$, the change in longitude $\delta \phi$ can be computed through the relation:
\begin{align} \label{eq:rt2}
\cos (vt) = \sin r_0 \sin r +  \cos r_0 \cos r' \cos( \delta \phi).
\end{align}
The set of kinematic equations \ref{eq:ta_si} and \ref{eq:rt2} are used to achieve a RTW on the surface of the sphere.
\end{document}